\documentclass[aps,prl,onecolumn,superscriptaddress,amsmath,amssymb]{revtex4}
\usepackage{graphicx}
\usepackage{txfonts}
\usepackage{amsfonts}
\usepackage{setspace}
\usepackage{graphicx}     % Include figure files
\usepackage{bm}
\usepackage{float}
\usepackage{pstricks}
\usepackage{pst-node}
\usepackage{pst-blur}
\restylefloat{table}
\definecolor{Pink}{rgb}{1.,0.75,0.8}

\begin{document}

%\preprint{}
\title{A magnetic origin for high temperature superconductivity in iron pnictides}

\author{Meng Wang$^\ast$}
\affiliation{Beijing National Laboratory for Condensed Matter Physics,
Institute of Physics, Chinese Academy of Sciences, Beijing 100190, China
}
\author{Chenglin Zhang$^\ast$}
\affiliation{
Department of Physics and Astronomy, The University of Tennessee, Knoxville, Tennessee 37996-1200, USA
}
\author{Xingye Lu$^\ast$}
\affiliation{Beijing National Laboratory for Condensed Matter Physics,
Institute of Physics, Chinese Academy of Sciences, Beijing 100190, China
}
\affiliation{
Department of Physics and Astronomy, The University of Tennessee, Knoxville, Tennessee 37996-1200, USA
}
\author{Guotai Tan}
\affiliation{
Department of Physics and Astronomy, The University of Tennessee, Knoxville, Tennessee 37996-1200, USA
}
\author{Huiqian Luo}
\affiliation{Beijing National Laboratory for Condensed Matter Physics,
Institute of Physics, Chinese Academy of Sciences, Beijing 100190, China
}
\author{Yu Song}
\affiliation{
Department of Physics and Astronomy, The University of Tennessee, Knoxville, Tennessee 37996-1200, USA
}
\author{Miaoyin Wang}
\affiliation{
Department of Physics and Astronomy, The University of Tennessee, Knoxville, Tennessee 37996-1200, USA
}
\author{Xiaotian Zhang}
\affiliation{Beijing National Laboratory for Condensed Matter Physics,
Institute of Physics, Chinese Academy of Sciences, Beijing 100190, China
}
\author{E. A. Goremychkin}
\affiliation{
ISIS Facility, Rutherford Appleton Laboratory, Chilton, Didcot, Oxfordshire OX11 0QX, UK
}
\author{T. G. Perring}
\affiliation{
ISIS Facility, Rutherford Appleton Laboratory, Chilton, Didcot, Oxfordshire OX11 0QX, UK
}
\author{T. A. Maier}
\affiliation{
Center for Nanophase Materials Sciences and Computer Science and Mathematics Division,
Oak Ridge National Laboratory, Oak Ridge, Tennessee 37831-6494, USA
}
\author{Zhiping Yin}
\affiliation{
Department of Physics, Rutgers University, Piscataway, NJ 08854, USA
}
\author{Kristjan Haule}
\affiliation{
Department of Physics, Rutgers University, Piscataway, NJ 08854, USA
}
\author{Gabriel Kotliar}
\affiliation{
Department of Physics, Rutgers University, Piscataway, NJ 08854, USA
}
\author{Pengcheng Dai}
\affiliation{
Department of Physics and Astronomy, The University of Tennessee, Knoxville, Tennessee 37996-1200, USA
}
\affiliation{Beijing National Laboratory for Condensed Matter Physics,
Institute of Physics, Chinese Academy of Sciences, Beijing 100190, China
}


\maketitle
{\bf In conventional Bardeen-Cooper-Schrieffer (BCS) superconductors \cite{bcs}, superconductivity occurs when electrons form coherent Cooper pairs
below the superconducting transition temperature $T_c$.
Although the kinetic energy of paired electrons increases in the superconducting state relative to the normal state, the reduction in the ion lattice energy is sufficient to give
the superconducting condensation energy ($E_c=-N(0)\Delta^2/2$ and $\Delta\approx 2\hbar\omega_D e^{-1/N(0)V_0}$,
where $N(0)$ is the electron density of states at zero temperature, $\hbar\omega_D$ is
the Debye energy, and $V_0$ is the strength electron-lattice coupling) \cite{chester,scalapino98,scalapino}.
For iron pnictide superconductors derived from electron or hole doping of
their antiferromagnetic (AF) parent compounds \cite{kamihara,rotter,mrotter,cruz,paglione10,dai},
the microscopic origin for superconductivity is unclear \cite{hirschfeld}. Here we use neutron scattering to show that high-$T_c$ superconductivity only occurs for iron pnictides with
low-energy ($\leq 25$ meV or $\sim 6.5k_BT_c$) itinerant electron-spin excitation coupling and
high energy ($>100$ meV) spin excitations.
Since our absolute spin susceptibility measurements for optimally hole-doped iron pnictide reveal that the change in magnetic exchange energy
below and above $T_c$ can
account for the superconducting condensation energy, we conclude that the presence of both
high-energy spin excitations giving rise to a large magnetic
exchange coupling $J$ and
low-energy spin excitations coupled to the itinerant electrons is essential
for high-$T_c$ superconductivity in iron pnictides.
 }

For BCS superconductors, the superconducting condensation energy $E_c$ and $T_c$ are controlled by the strength of
the Debye energy $\hbar\omega_D$ and electron-lattice coupling $V_0$ \cite{bcs}.  A material with large $\hbar\omega_D$ and lattice exchange coupling
is a necessary but not a sufficient condition to have high-$T_c$ superconductivity.  On the other hand, a soft metal with small $\hbar\omega_D$ (such as lead and mercury)
will also not exhibit superconductivity with high-$T_c$.  For
unconventional superconductors such as iron pnictides, the superconducting phase is derived from hole and electron doping from
their AF parent compounds \cite{kamihara,rotter,mrotter,cruz,paglione10,dai}.
Although the static long-range AF order
is gradually suppressed when electrons or holes are doped into the iron pnictide parent compound
such as BaFe$_2$As$_2$ \cite{rotter,mrotter,cruz,paglione10,dai}, short-range spin excitations remain throughout the superconducting phase and are coupled directly
with the occurrence of superconductivity \cite{christianson,castellan,chenglinzhang,chlee11,lumsden,schi09,inosov,jtpark,mliu,huiqian}.  For spin excitations
mediated superconductors,
the superconducting condensation energy should be accounted for by the change in magnetic exchange energy between the normal ($N$) and superconducting ($S$)
phases at zero temperature.  Within the $t$-$J$ model \cite{spalek},
$\Delta E_{ex}(T)=2J[\left\langle {\bf S}_{i+x}\cdot{\bf S}_i\right\rangle_N-
\left\langle {\bf S}_{i+x}\cdot{\bf S}_i\right\rangle_S]$, where $\left\langle {\bf S}_{i+x}\cdot{\bf S}_i\right\rangle$ is the dynamic spin
susceptibility in absolute units at temperature $T$ \cite{scalapino98,scalapino}.

To determine how high-$T_c$ superconductivity in iron pnictides is associated with
spin excitations, we consider the phase diagram of electron and hole-doped iron pnictide BaFe$_2$As$_2$
(Fig. 1a) \cite{dai}.  In the undoped state, BaFe$_2$As$_2$ forms a
metallic low-temperature orthorhombic phase with collinear AF structure as shown
in the inset of Fig. 1a.  Inelastic neutron scattering (INS) measurements
have mapped out spin waves throughout the Brillouin zone, and determined the effective magnetic exchange couplings  \cite{harriger}.
Upon doping electrons to BaFe$_2$As$_2$ by partially replacing Fe with Ni to induce
superconductivity in BaFe$_{2-x}$Ni$_x$As$_2$ with maximum $T_c\approx20$ K at $x_e=0.1$ \cite{ljli}, the low-energy
($<80$ meV) spin waves in the parent compounds are broadened and
form a neutron spin resonance coupled to superconductivity \cite{lumsden,schi09,inosov,jtpark}, while high-energy spin excitations remain unchanged \cite{mliu}.
With further electron-doping to $x_e\ge0.25$, superconductivity is suppressed and the system
becomes a paramagnetic metal (Fig. 1a) \cite{ljli}.
For hole-doped Ba$_{1-x}$K$_x$Fe$_2$As$_2$ \cite{savci}, superconductivity with maximum $T_c=38.5$ K appears
at $x_h\approx 0.33$ \cite{rotter} and pure KFe$_2$As$_2$ at $x_h=1$ is a $T_c=3.1$ K superconductor \cite{mrotter}.
In order to determine how spin excitations throughout the Brillouin zone
are correlated with superconductivity
in iron pnictides, we study optimally hole-doped
Ba$_{0.67}$K$_{0.33}$Fe$_2$As$_2$ ($T_c=38.5$ K, Fig. 1c), pure KFe$_2$As$_2$ ($T_c=3$ K, Fig. 1b), and nonsuperconducting
electron-overdoped BaFe$_{1.7}$Ni$_{0.3}$As$_2$ (Fig. 1d).
If spin excitations
are responsible for mediating electron pairing and superconductivity, the change
in magnetic exchange energy between the normal ($N$) and superconducting ($S$) state
should be large enough to account for the superconducting
condensation energy \cite{scalapino}.

\begin{figure}[t]
\includegraphics[scale=.6]{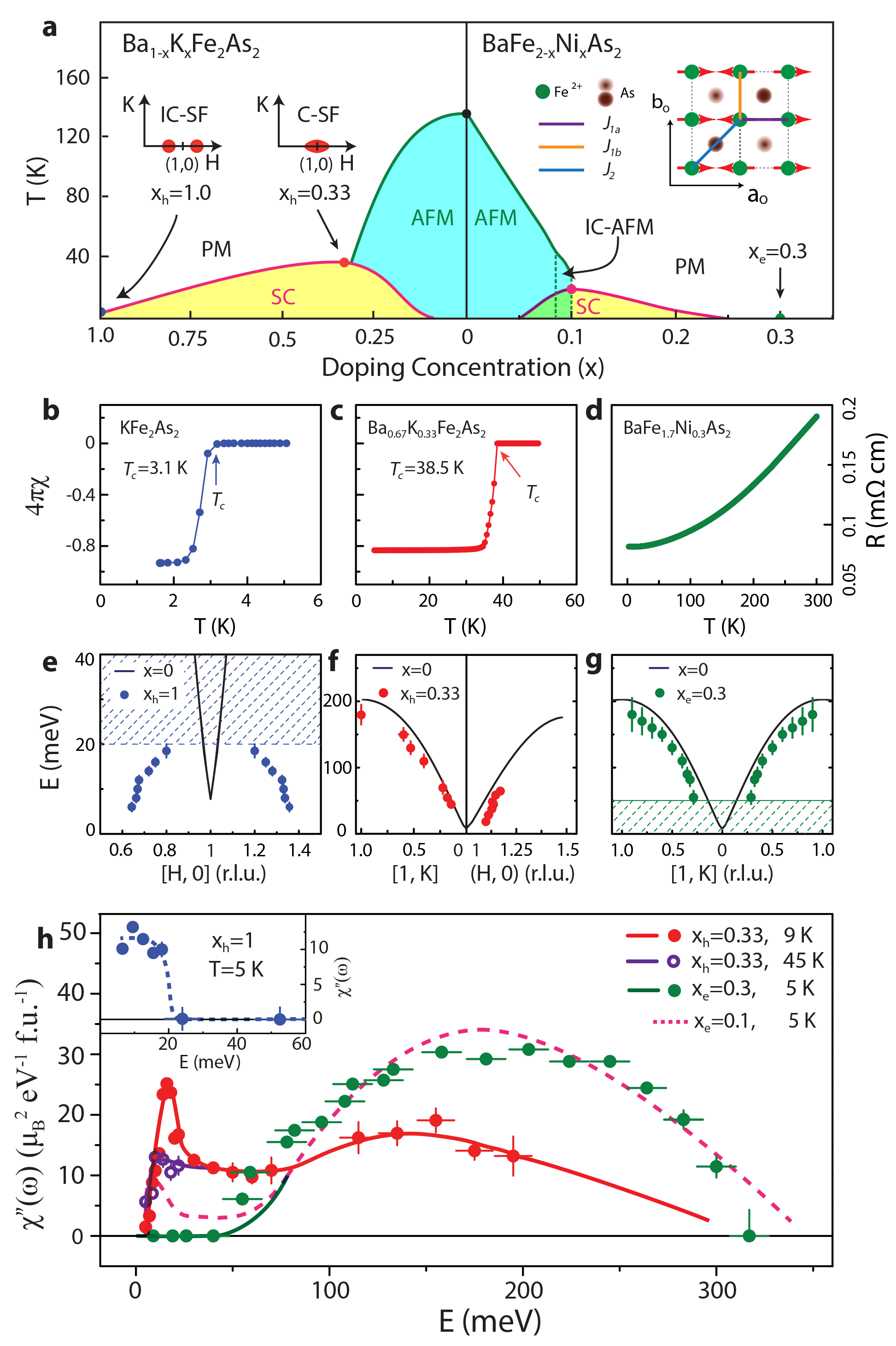}
\caption{ {\bf Summary of transport, magnetic, and neutron scattering results}.
Our experiments were carried out on the MERLIN and MAPS time-of-flight chopper spectrometers at the Rutherford-Appleton Laboratory, UK \cite{mliu,harriger}.
We co-aligned 19 g of single crystals of Ba$_{0.67}$K$_{0.33}$Fe$_2$As$_2$ (with in-plane and out-of-plane
mosaic of $\sim$4$^\circ$), 3 g of KFe$_2$As$_2$ (with in-plane and out-of-plane
mosaic of $\sim$7.5$^\circ$), and  40 g of BaFe$_{1.7}$Ni$_{0.3}$As$_2$ (with in-plane and out-of-plane
mosaic of $\sim$3$^\circ$). Various incident beam energies were used as specified,
and mostly with $E_i$ parallel to the $c$-axis.  To facilitate easy comparison with spin waves in BaFe$_{2}$As$_{2}$ \cite{harriger}, we
defined the wave vector $Q$ at ($q_x$, $q_y$, $q_z$) as $(H,K,L) = (q_xa/2\pi, q_yb/2\pi, q_zc/2\pi)$ reciprocal lattice units (rlu) using the
orthorhombic unit cell, where $a = b = 5.57$ \AA, and $c = 13.135$ \AA\ for Ba$_{0.67}$K$_{0.33}$Fe$_2$As$_2$,
$a = b = 5.43$ \AA, and $c = 13.8$ \AA\ for KFe$_2$As$_2$, and $a = b = 5.6$ \AA, and $c = 12.96$ \AA\ for BaFe$_{1.7}$Ni$_{0.3}$As$_2$.
The data are normalized to absolute units using a vanadium standard with 20\% errors \cite{mliu}.
(a) The electronic phase diagram of electron and hole-doped BaFe$_2$As$_2$ \cite{dai}.  The right inset shows
crystal and AF spin structures of
 BaFe$_{2}$As$_{2}$ with marked the nearest ($J_{1a}$, $J_{1b}$) and next nearest neighbor ($J_2$) magnetic
 exchange couplings.  The left insets show the evolution of low-energy spin excitations in Ba$_{1-x}$K$_x$Fe$_2$As$_2$.
(b,c) Temperature dependence of magnetic susceptibility for our KFe$_2$As$_2$ and Ba$_{0.67}$K$_{0.33}$Fe$_2$As$_2$.
(d)  Temperature dependence of the resistivity for BaFe$_{1.7}$Ni$_{0.3}$As$_2$.
(e,f,g) The filled circles
are spin excitation dispersions of KFe$_2$As$_2$ at 5 K, Ba$_{0.67}$K$_{0.33}$Fe$_2$As$_2$ at 9 K, and BaFe$_{1.7}$Ni$_{0.3}$As$_2$ at 5 K, respectively.
The shaded areas indicate vanishing spin excitations and the solid lines show spin wave dispersions of BaFe$_2$As$_2$ \cite{harriger}.
(h) Energy dependence of $\chi^{\prime\prime}(\omega)$ for
BaFe$_{1.9}$Ni$_{0.1}$As$_2$ (dashed line), BaFe$_{1.7}$Ni$_{0.3}$As$_2$ (green solid circles),
Ba$_{0.67}$K$_{0.33}$Fe$_2$As$_2$ below (Solid red circles and solid red line)
and above (open purple circles and solid lines) $T_c$.  The inset shows Energy dependence of $\chi^{\prime\prime}(\omega)$ for KFe$_2$As$_2$.
 The vertical error
bars indicate the statistical errors of one standard deviation.  The horizontal error bars in (h) indicate the energy integration range.
 }
\end{figure}

In previous INS work
on powder \cite{christianson,castellan} and single crystals \cite{chenglinzhang} of hole-doped Ba$_{1-x}$K$_{x}$Fe$_2$As$_2$, low-energy spin excitations were found to be dominated by a resonance coupled to superconductivity.  In fact, density functional theory (DFT) calculations based
on sign reversed quasiparticle excitations between hole Fermi surface pocket near the Brillouin zone center and electron pocket near the zone corner (see supplementary information)
\cite{hirschfeld} predict correctly the longitudinally
elongated spin excitations from $Q_{AF}=(1,0)$ for optimally hole-doped iron pnictide (inset in Fig. 1a) \cite{jtpark,chenglinzhang}.
For pure KFe$_2$As$_2$ ($x_h=1$), low-energy ($<14$ meV) spin excitations become longitudinally incommensurate from $Q_{AF}$ (inset in Fig. 1a) \cite{chlee11}.
These results, as well as work on electron-doped iron pnictides that reveal transversely elongated spin excitations from $Q_{AF}$ \cite{jtpark,mliu,huiqian}, have shown that the low energy nematic-like spin dispersion in iron-based superconductors can be accounted for
by itinerant electrons on hole and electron nested Fermi surfaces \cite{dai}.

\begin{figure}[t]
\includegraphics[scale=.7]{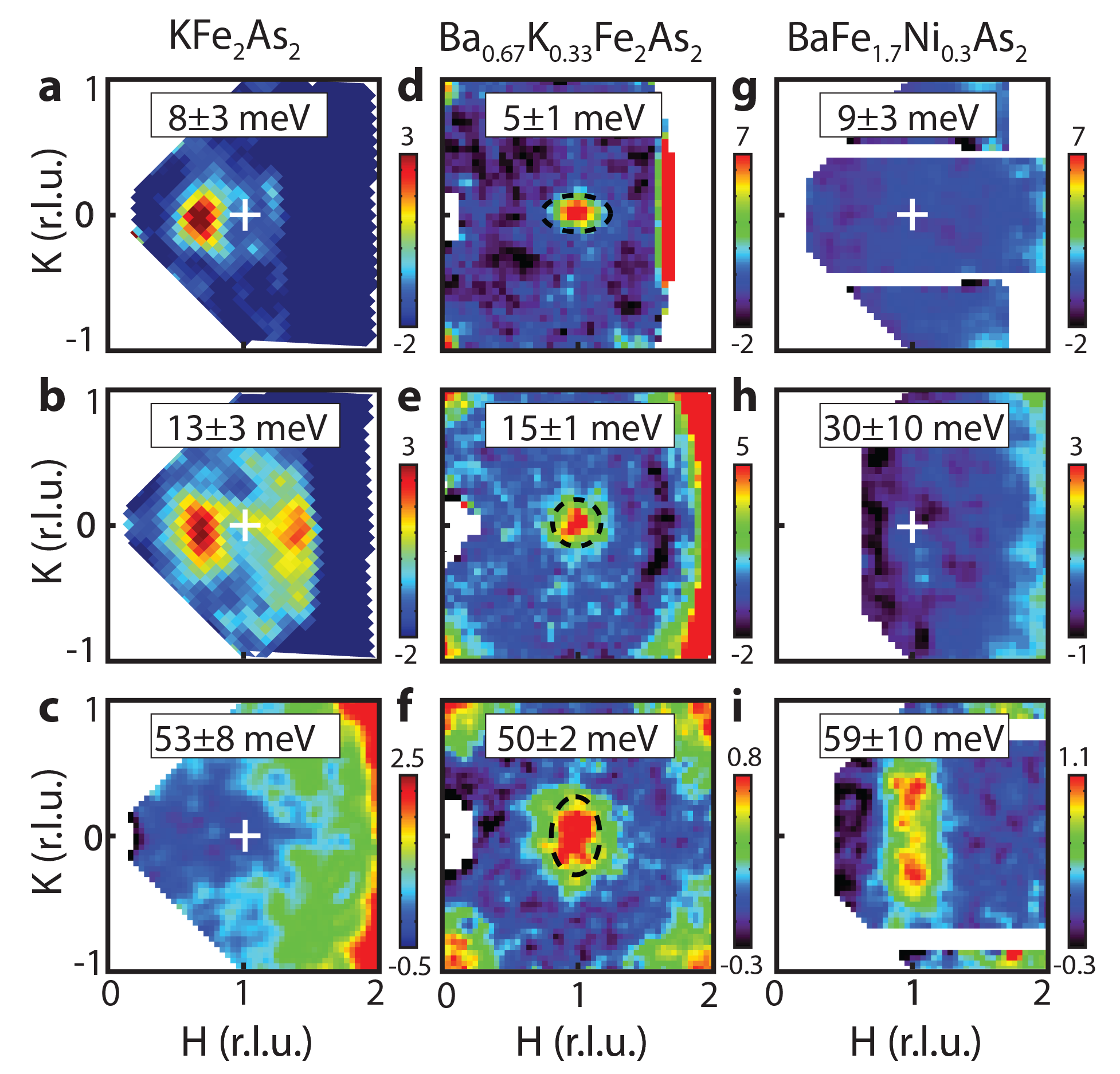}
\caption{
{\bf Constant-energy slices through magnetic excitations of KFe$_2$As$_2$,
Ba$_{0.67}$K$_{0.33}$Fe$_2$As$_2$, and BaFe$_{1.7}$Ni$_{0.3}$As$_2$ at different energies}.
The color bars represent the vanadium normalized absolute spin excitation intensity in the units of
mbarn/sr/meV/f.u. Two dimensional images of spin excitations at 5 K for KFe$_2$As$_2$
(a) $E=8\pm 3$ meV obtained with $E_i=20$ meV. The right side incommensurate peak is obscured by background scattering.
(b) $13\pm 3$ meV with $E_i=35$ meV, (c) $53\pm 10$ meV with
$E_i=80$ meV.  For Ba$_{0.67}$K$_{0.33}$Fe$_2$As$_2$ at $T=45$ K, images of spin excitations at
(d) $E=5\pm 1$ meV obtained with $E_i=20$ meV, (e) $15\pm 1$ meV with $E_i=35$ meV, and
(f) $50\pm 2$ meV
obtained with $E_i=80$ meV.
The dashed box in (d) indicates the AF zone boundaries for a single FeAs layer and the black dashed
lines mark the orientations of spin excitations at different energies.
Images of spin excitations for BaFe$_{1.7}$Ni$_{0.3}$As$_2$ at $T=5$ K and
(g) $E=9\pm3$ meV obtained with $E_i=80$ meV, (h) $30\pm10$ meV with $E_i=450$ meV, and
(i) $59\pm10$ meV with $E_i=250$ meV.  The white crosses indicate the position of $Q_{AF}$.
 }
\end{figure}

Here we use INS to show that
the effective magnetic exchange couplings in hole-doped Ba$_{0.67}$K$_{0.33}$Fe$_2$As$_2$
only soften slightly from that of AF BaFe$_2$As$_2$ \cite{harriger}
and electron-doped BaFe$_{1.9}$Ni$_{0.1}$As$_{2}$ superconductor (Fig. 1f) \cite{mliu}.  The effect of hole-doping in BaFe$_2$As$_2$ is to suppress
high-energy spin excitations and transfer
the spectral weight to low-energies that couple to the appearance of superconductivity (Fig. 1h).
This is qualitatively consistent with theoretical methods based on DFT and dynamic mean filed theory (DMFT, see supplementary information) \cite{park}.
By using the INS measured magnetic exchange couplings and spin susceptibility
in absolute units, we calculate the superconductivity-induced lowering of
magnetic exchange energy and find it to be about seven times larger than
the superconducting condensation energy determined from specific heat measurements for Ba$_{0.67}$K$_{0.33}$Fe$_2$As$_2$ \cite{popovich}.
These results are consistent with spin excitations mediated electron pairing mechanism \cite{scalapino}.
For the nonsuperconducting electron overdoped BaFe$_{1.7}$Ni$_{0.3}$As$_{2}$, we find
that while the effective magnetic exchange couplings only reduce slightly compared with that
of optimally electron-doped BaFe$_{1.9}$Ni$_{0.1}$As$_{2}$ (Fig. 1g) \cite{mliu},
the low-energy spin excitations ($<50$ meV) associated with hole and electron pocket Fermi surface nesting disappear, thus revealing
the importance of Fermi surface nesting and itinerant electron-spin excitation coupling to the  occurrence of superconductivity (Fig. 1h).
Finally, for heavily hole-doped KFe$_2$As$_2$ with low-$T_c$ superconductivity (Fig. 1b), there are only incommensurate spin
excitations below $\sim$25 meV and
the correlated
high-energy spin excitations prevalent in electron-doped and optimally hole-doped iron pnictides are completely suppressed (Fig. 1e), indicating a dramatic softening
of effective magnetic exchange coupling (inset Fig. 1h).  Therefore,
high-$T_c$ superconductivity in iron pnictides requires two fundamental ingredients: a
large effective magnetic exchange coupling \cite{scalapino}, much like
large Debye energy for high-$T_c$ BCS superconductors, and a strong itinerant electrons-spin excitations coupling
from Fermi surface nesting \cite{hirschfeld}, like electron-phonon coupling in BCS superconductors.  The presence of correlated electronic states
exhibiting both local and itinerant properties is essential
for the mechanism of superconductivity \cite{dai}.

\begin{figure}[t]
\includegraphics[scale=.7]{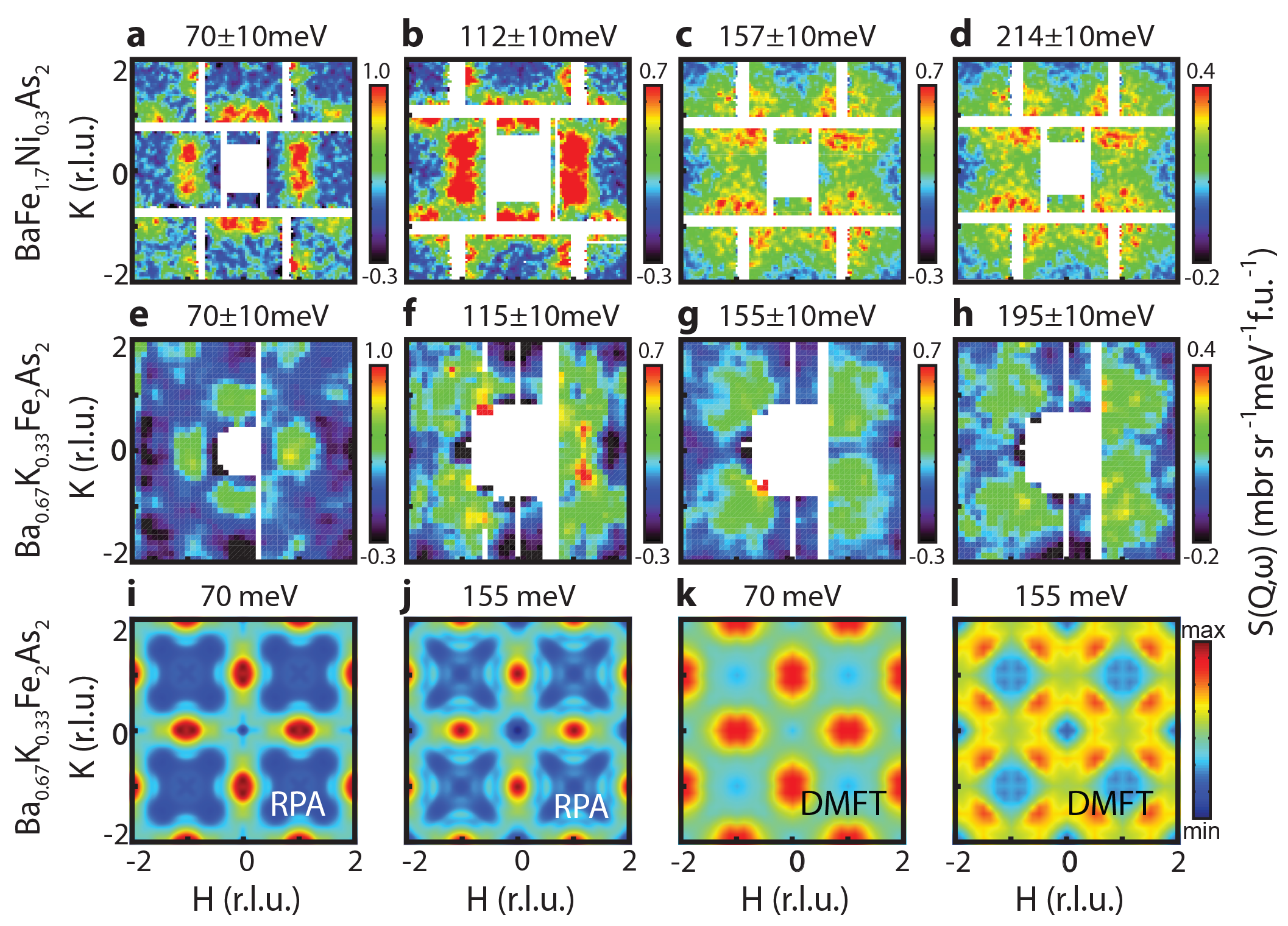}
\caption{
{\bf Constant-energy images of spin excitations in the 2D $[H,K]$ plane at different energies for BaFe$_{1.7}$Ni$_{0.3}$As$_2$ and
Ba$_{0.67}$K$_{0.33}$Fe$_2$As$_2$ and its comparison with RPA/DMFT calculations for Ba$_{0.67}$K$_{0.33}$Fe$_2$As$_2$}.
Spin excitations of BaFe$_{1.7}$Ni$_{0.3}$As$_2$ at energy transfers of (a)
$E=70\pm10$ meV obtained with $E_i=250$ meV;(b) $112\pm 10$ meV  $E_i=250$ meV;(c) $157\pm10$ meV; (d) $214\pm10$ meV, all obtained with $E_i=450$ meV
at 5 K. A flat backgrounds have been subtracted from the images.
Spin excitations of Ba$_{0.67}$K$_{0.33}$Fe$_2$As$_2$ at energy transfers of
(e) $E=70\pm10$ meV obtained with $E_i=170$ meV;(f) $115\pm 10$ meV;(g) $155\pm10$ meV; (h) $195\pm10$ meV, all obtained with $E_i=450$ meV at 9 K.
Wave vector dependent backgrounds have been subtracted from the images.
RPA calculations \cite{huiqian} of spin excitations for Ba$_{0.67}$K$_{0.33}$Fe$_2$As$_2$ at (i) $E=70$ meV and (j) $E=155$ meV.
DMFT calculations \cite{mliu,park} for Ba$_{0.67}$K$_{0.33}$Fe$_2$As$_2$ at (k) $E=70$ meV and (l) $E=155$ meV.
 }
\end{figure}

To substantiate the key conclusions of Fig. 1, we present the two-dimensional (2D) constant-energy images of
spin excitations in the
$(H, K)$ plane at different energies for KFe$_2$As$_2$ (Figs. 2a-2c), Ba$_{0.67}$K$_{0.33}$Fe$_2$As$_2$ (Figs. 2d-2f), and
BaFe$_{1.7}$Ni$_{0.3}$As$_2$ (Figs. 2g-2i) above $T_c$.
In previous INS work on KFe$_2$As$_2$, longitudinal incommensurate spin excitations were found by triple axis spectrometer measurements
for energies from 3 to 14 meV in the normal state \cite{chlee11}.  While we confirmed the earlier work using time-of-flight INS
for energies below $E=15\pm 1$ meV (Figs. 2a and 2b), our new data collected at higher excitation energies reveals that
incommensurate spin excitations converge into a broad spin excitation near $E=20$ meV and disappear for energies above
25 meV (Fig. 2c).  Therefore, there are no measurable correlated spin excitations for energies above 30 meV, indicating that
the effective magnetic coupling in KFe$_2$As$_2$ has reduced about 90\% compared with that of BaFe$_2$As$_2$ (see supplementary information).
For Ba$_{0.67}$K$_{0.33}$Fe$_2$As$_2$, spin excitations at
$E = 5\pm 1$ meV are longitudinally elongated from $Q_{AF}$
as expected from DFT calculations (Fig. 2d) \cite{jtpark,chenglinzhang}.  At the resonance energy  ($E=15\pm 1$ meV) \cite{christianson},
spin excitations are isotropic above $T_c$ (Fig. 2e).
On increasing energy further to $E=50\pm 10$ meV, spin excitations change to transversely elongated
from $Q_{AF}$ similar to spin excitations in optimally electron-doped superconductor BaFe$_{1.9}$Ni$_{0.1}$As (Fig. 2f) \cite{mliu}.
Figures 2g-2i summarize similar 2D constant-energy images of
spin excitations for nonsuperconducting BaFe$_{1.7}$Ni$_{0.3}$As.  At $E=9\pm3$ (Fig. 2g) and $30\pm 10$ meV (Fig. 2h), there are no correlated
spin excitations near the $Q_{AF}$.  Upon increasing energy to $E=59\pm 10$ meV (Fig. 2i), we see clear spin excitations
transversely elongated from $Q_{AF}$ (Fig. 2i).

Figures 3a-3d show 2D images of spin excitations in BaFe$_{1.7}$Ni$_{0.3}$As$_2$ at
$E=70\pm 10$, $112\pm 10$, $157\pm 10$, and $214\pm 10$ meV, respectively.
Figures 3e-3h show wave vector dependence of spin excitations at energies
$E=70\pm 10, 115\pm10, 155\pm 10$, and $195\pm10$ meV, respectively, for Ba$_{0.67}$K$_{0.33}$Fe$_2$As$_2$.
Similar to spin waves in BaFe$_2$As$_2$ \cite{harriger}, spin excitations in BaFe$_{1.7}$Ni$_{0.3}$As$_2$ and
Ba$_{0.67}$K$_{0.33}$Fe$_2$As$_2$
split along the $K$-direction for energies above 80 meV and form rings around $Q=(\pm1,\pm1)$ positions
near the zone boundary, albeit at slightly different energies.
Comparing spin excitations in Figs. 3a-3d for BaFe$_{1.7}$Ni$_{0.3}$As$_2$
with those in Figs. 3e-3h for Ba$_{0.67}$K$_{0.33}$Fe$_2$As$_2$ in absolute units,
we see that spin excitations
in BaFe$_{1.7}$Ni$_{0.3}$As$_2$ extend to slightly higher energies and have larger intensity above 100 meV.

To understand the wave vector dependence of spin excitations in
hole-doped Ba$_{0.67}$K$_{0.33}$Fe$_2$As$_2$, we have carried out
the random
phase approximation (RPA) calculation of
the dynamic susceptibility
in a pure itinerant electron picture using method described before \cite{huiqian}.
Figures 3i and 3j show RPA calculations of spin excitations at $E=70$ and 155 meV, respectively, for
Ba$_{0.67}$K$_{0.33}$Fe$_2$As$_2$ assuming that hole doping induces a
rigid band shift \cite{huiqian}.  The outcome is in clear disagreement with Figs. 3e and 3g, indicating
that a pure RPA type itinerant model cannot describe the wave vector dependence of spin excitations in hole-doped iron pinctides
at high energies.
For comparison with the RPA calculation, we also used a combined DFT and DMFT approach \cite{mliu,park}
 to calculate the imaginary part of the dynamic susceptibility
$\chi^{\prime\prime}({\bf Q},\omega)$ in the paramagnetic state.  Figures
3k and 3i show calculated spin excitations at $E=70$ and 155 meV, respectively.  Although the model still does not agree in detail
with the data in Figs. 3e and 3g, it captures the trend of spectral weight transfer away from
$Q_{AF}=(1,0)$ on increasing the energy and forming a pocket centered at $Q=(1,1)$.

 By carrying out cuts through
the 2D images similar to Figs. 2d-2f and 3e-3h
along the $[1,K]$ and $[H,0]$ directions (see supplementary information),
we establish the spin excitation dispersion along the
two high symmetry directions for Ba$_{0.67}$K$_{0.33}$Fe$_2$As$_2$ and compare with the dispersion of BaFe$_2$As$_2$
(Fig. 1f) \cite{harriger}.  In contrast to the dispersion of electron-doped BaFe$_{1.9}$Ni$_{0.1}$As$_{2}$ \cite{mliu}, we find clear softening of the zone boundary spin excitations in hole-doped Ba$_{0.67}$K$_{0.33}$Fe$_2$As$_2$ from spin waves in BaFe$_2$As$_2$ \cite{harriger}.  We estimate that the effective magnetic exchange coupling in Ba$_{0.67}$K$_{0.33}$Fe$_2$As$_2$
is reduced by about 10\% from that of BaFe$_2$As$_2$ (see supplementary information).
Similarly,  Figure 1g shows the dispersion curve of
BaFe$_{1.7}$Ni$_{0.3}$As$_2$ along the $[1,K]$ direction plotted together with that of BaFe$_2$As$_2$
\cite{harriger}.  For energies below $\sim$50 meV, spin excitations are completely gapped marked
in the dashed area probably due to the
missing hole-electron Fermi pocket quasiparticle excitations
\cite{hirschfeld,prichard}.
Based on the 2D spin excitation images similar to Figs. 2a-2c, we plot in Fig. 1e the dispersion of incommensurate spin excitations in KFe$_2$As$_2$.
The incommensurability of spin excitations is weakly energy dependent below $E=12$ meV but becomes smaller with
increasing energy above 12 meV (see supplementary information).  Correlated spin excitations for energies above $25$ meV are
completely suppressed as shown in the shaded area in Fig. 1e.

\begin{figure}[t]
\includegraphics[scale=.6]{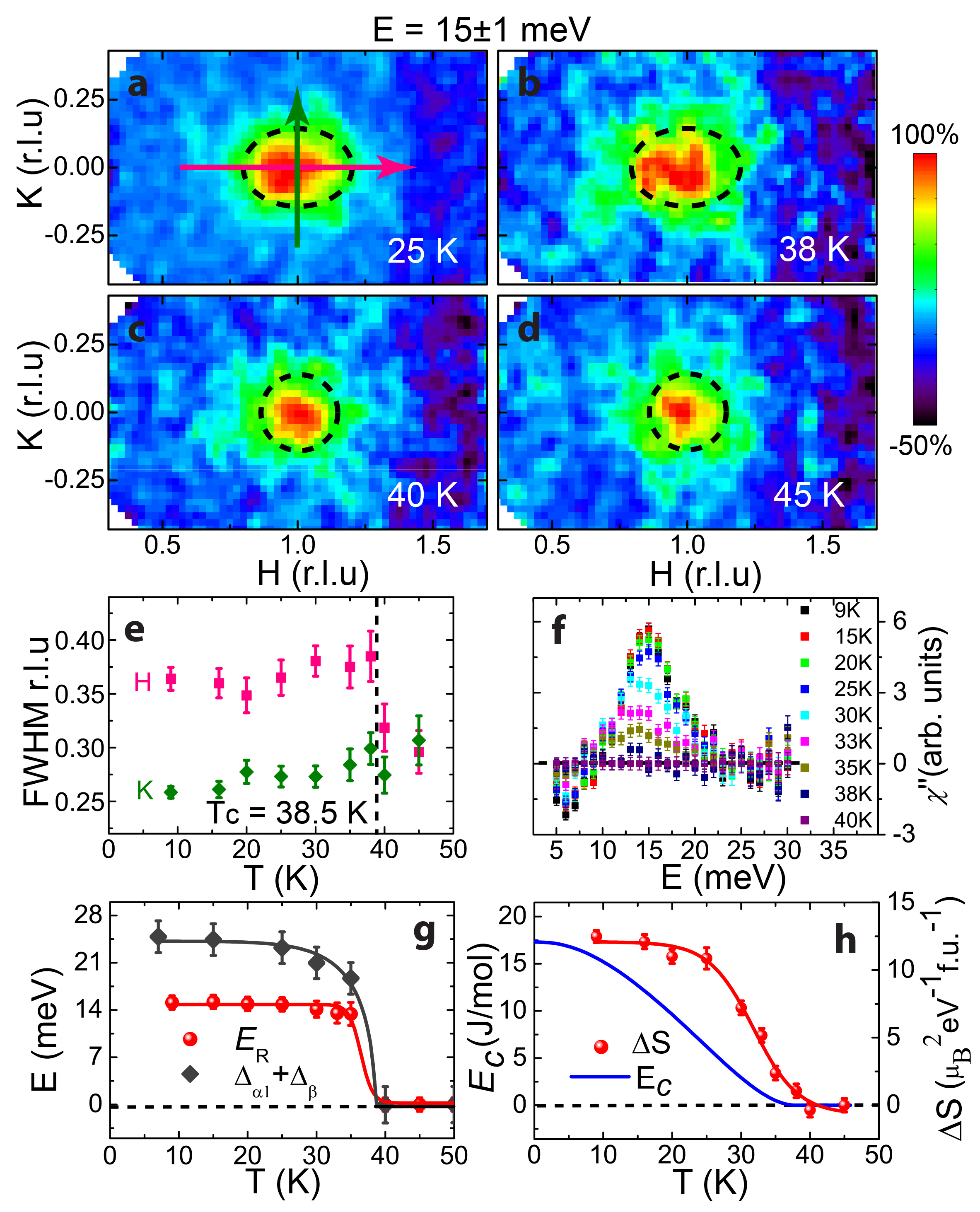}
\caption{ {\bf Wave vector, energy, and temperature dependence of the resonance across $T_c=38.5$ K}.
Constant-energy ($E=15\pm 1$ meV) images of spin excitations at (a) $T=25$, (b) 38, (c) 40, and (d)45 K obtained with
$E_i=35$ meV. In order to make fair comparison of the scattering line shape at different temperatures,
the peak intensity at each temperature is normalized to 1.  The pink and green arrows in (a) mark
wave vector cut directions across the resonance. The integration ranges are $-0.2\leq K\leq 0.2$ along the $[H,0]$ direction
and $0.8\leq H\leq 1.2$ along the $[1,K]$ direction.  The FWHM of spin excitations are marked as dashed lines.
(e) The FWHM of the resonance
along the $[H,0]$ and $[1,K]$ directions as a function of temperature across $T_c$. (f) Energy dependence of the
resonance obtained by subtracting the low-temperature data from the 45 K data, and
correcting for the Bose population factor.  (g) The black diamonds show temperature dependence of
the sum of hole and electron pocket electronic gaps obtained from Angle Resolved
Photoemission experiments for Ba$_{0.67}$K$_{0.33}$Fe$_2$As$_2$ \cite{prichard}.  The red solid circles show
temperature dependence of the resonance. (h) Temperature
dependence of the superconducting condensation energy from heat capacity measurements \cite{popovich} and
the intensity of the resonance integrated from 14 to 16 meV.
 The error bars indicate the statistical errors of one standard deviation.
 }
\end{figure}

To quantitatively determine the effect of electron and hole doping on
the overall spin excitations spectra, we
calculate the local dynamic susceptibility per formula unit
(f.u.) in absolute units, defined as
$\chi^{\prime\prime}(\omega)=\int{\chi^{\prime\prime}({\bf
    q},\omega)d{\bf q}}/\int{d{\bf q}}$ \cite{mliu},
where $\chi^{\prime\prime}({\bf q},\omega)=(1/3) tr( \chi_{\alpha \beta}^{\prime\prime}({\bf q},\omega))$,
at different energies for Ba$_{0.67}$K$_{0.33}$Fe$_2$As$_2$, BaFe$_{1.7}$Ni$_{0.3}$As$_2$, and
KFe$_2$As$_2$.  Figure 1h shows the outcome together with previous data on
optimally electron-doped superconductor BaFe$_{1.9}$Ni$_{0.1}$As$_{2}$ \cite{mliu}.
While electron-doping up to BaFe$_{1.7}$Ni$_{0.3}$As$_2$ does not change the spectral weight of high-energy
spin excitations from that of BaFe$_{1.9}$Ni$_{0.1}$As$_{2}$, hole-doping dramatically suppresses the high-energy spin excitations
and shift the spectral weight to lower energies (Fig. 1h).  For heavily hole-doped KFe$_2$As$_2$, spin excitations are confined to energies below about $E=25$ meV (inset in Fig. 1h).
The reduction of the high energy spin spectral weight and its transfer to low energy with hole doping, but not with electron doping, is not naturally explained by the band theory,  and requires models which incorporate both the itinerant quasiparticles and the local moment physics (see supplementary information) \cite{park}.
The hole doping makes electronic state more correlated, as local moment formation is strongest in the half-filled $d^5$ shell,
and mass enhancement larger thereby reducing the electronic energy scale in the problem.

Finally, to determine how low-energy spin excitations are
coupled to superconductivity in Ba$_{0.67}$K$_{0.33}$Fe$_2$As$_2$, we carried out a detailed temperature dependent study of spin excitation at $E=15\pm 1$ meV. Comparing with strongly $c$-axis modulated low-energy ($E<7$ meV) spin excitations \cite{chenglinzhang}, spin excitations at the resonance energy
are essentially 2D without much $c$-axis modulations.
Figures 4a-4d show the 2D mapping of the resonance at $T=25, 38, 40,$ and 45 K, respectively.
While the resonance reveals a clear oval shape at temperatures below $T_c$ (Figs. 4a and 4b),
it changes into an isotropic circular
shape abruptly at $T_c$ (Figs. 4c and 4d) as shown by the dashed lines
representing full-width-at-half-maximum (FWHM) of the excitations.   Temperature dependence of the resonance width
along the $[H,0]$ and $[1,K]$ directions in Fig. 4e
reveals that the isotropic to anisotropic transition in momentum space occurs at
$T_c$.
Figure 4f shows temperature dependence of the resonance from 9 K to 40 K, which vanishes at $T_c$.
Figure 4g plots temperature dependence of the mode energy together with the sum of the
superconducting gaps from hole and electron pockets \cite{prichard}.
Figure 4h compares temperature dependence of the superconducting condensation energy \cite{popovich} with superconductivity
induced intensity gain of the resonance.  By calculating
spin excitations induced changes in magnetic exchange energy (see supplementary information) \cite{scalapino},
we find that the difference of magnetic exchange interaction
energy between the superconducting and normal state is approximately
seven times larger than
the superconducting condensation energy \cite{popovich}, thus identifying
AF spin excitations as the major driving force for superconductivity in Ba$_{0.67}$K$_{0.33}$Fe$_2$As$_2$.

One way to quantitatively estimate the impact of hole/electron doping and superconductivity
to spin waves of BaFe$_2$As$_2$ is to determine the energy dependence of
the local moment and total fluctuating moments $\left\langle m^2\right\rangle$ \cite{mliu}.
From Fig. 1h, we see that hole-doping suppresses high-energy spin waves of BaFe$_2$As$_2$ and pushes the spectral weight to resonance
at lower energies.  The total fluctuating moment of Ba$_{0.67}$K$_{0.33}$Fe$_2$As$_2$ below 300 meV
is $\left\langle m^2\right\rangle=1.7\pm 0.3$ per Fe, somewhat smaller than
 $3.2\pm 0.16\ \mu_B^2$ per Fe
for BaFe$_2$As$_2$ and BaFe$_{1.9}$Ni$_{0.1}$As$_{2}$ \cite{mliu}.  For
comparison, BaFe$_{1.7}$Ni$_{0.3}$As$_{2}$ and KFe$_2$As$_2$ have $\left\langle m^2\right\rangle=2.74\pm 0.11$
and $0.1\pm0.02\ \mu_B^2$ per Fe, respectively.
Therefore, the total magnetic spectral weights for different iron pnictides have no direct correlation with
their superconducting $T_c$'s.
 From Fig. 1h, we also see that the spectral weight of the resonance
 and low-energy ($<100$ meV) magnetic scattering in Ba$_{0.67}$K$_{0.33}$Fe$_2$As$_2$ is much larger than that of electron-doped
BaFe$_{1.9}$Ni$_{0.1}$As$_{2}$.  This is consistent with a large superconducting condensation energy in
Ba$_{0.67}$K$_{0.33}$Fe$_2$As$_2$ since
its effective magnetic exchange coupling $J$
is only $\sim$10\% smaller than that of BaFe$_{1.9}$Ni$_{0.1}$As$_{2}$ (Fig. 1h) \cite{popovich,bzeng}.
For electron-overdoped nonsuperconducting BaFe$_{1.7}$Ni$_{0.3}$As$_{2}$,
the lack of superconductivity is
due to the absence of low-energy spin excitations coupled to the hole and electron Fermi surface nesting
even though the effective magnetic exchange remains large
\cite{hirschfeld,prichard}. This means that by eliminating $[\left\langle {\bf S}_{i+x}\cdot{\bf S}_i\right\rangle_N-
\left\langle {\bf S}_{i+x}\cdot{\bf S}_i\right\rangle_S]$, there is no magnetic driven superconducting condensation energy and thus
no superconductivity. On the other hand, although the complete suppression of correlated
high-energy spin excitations in KFe$_2$As$_2$ can dramatically reduce the effective magnetic exchange coupling in KFe$_2$As$_2$ (Fig. 1e),
one can still have superconductivity with reduced $T_c$.
If spin excitations are a common
thread of the electron pairing interactions for unconventional superconductors \cite{scalapino},
our results reveal that both the large effective magnetic exchange couplings and itinerant electron-spin excitation interactions
are essential ingredients to achieve high-$T_c$ superconductivity, much like large Debye energy and strength of electron-lattice coupling are necessary
for high-$T_c$ BCS superconductors.  Therefore,
the presence of correlated electronic states exhibiting wave-particle duality is important
for the mechanism of high-$T_c$ superconductivity in iron pnictides.

%\bibliography{NoEndingPoint}

\begin{flushleft}
{\bf Acknowledgements}
Work at IOP is supported
by the MOST of China 973 programs (2012CB821400, 2011CBA00110)
and NSFC-11004233.  The single crystal growth and neutron scattering
work at UTK is supported by the U.S. DOE
BES under Grant No. DE-FG02-05ER46202. The LDA+DMFT
computations were made possible by an Oak Ridge leadership computing facility  director discretion allocation to Rutgers.
The work at Rutgers is supported by DOE BES DE-FG02-99ER45761 (GK) and
NSF-DMR 0746395 (KH).
TAM acknowledges the Center for Nanophase Materials Sciences, which is sponsored at ORNL by the Scientific User Facilities Division, BES, U.S. DOE.
\end{flushleft}

\begin{flushleft}
{\bf Author contributions}
This paper contains data from three different neutron scattering experiments in the group of P. D.
lead by M.W. (Ba$_{0.67}$K$_{0.33}$Fe$_2$As$_2$), C.L.Z (KFe$_2$As$_2$), and X.Y.L (BaFe$_{1.7}$Ni$_{0.3}$As$_2$).
These authors made equal contributions to the results reported in the paper.
For Ba$_{0.67}$K$_{0.33}$Fe$_2$As$_2$, M.W., H.Q.L., E.A.G., and P.D. carried out neutron scattering experiments.
Data analysis was done by M.W. with help from H.Q.L., and E.A.G..  The samples were grown by C.L.Z., M.W., Y.S., X.Y.L., and co-aligned by M.W. and H.Q.L.
RPA calculation is carried out by T.A.M.
The DFT and DMFT calculations were done by Z.P.Y., K.H., and G.K.  Superconducting condensation energy was estimated by X.T.Z.
For KFe$_2$As$_2$, the samples were grown by C.L.Z and G.T.T. Neutron scattering experiments were carried out by C.L.Z., T.G.P., and P.D.
For BaFe$_{1.7}$Ni$_{0.3}$As$_2$, the samples were grown by X.Y.L., H.Q.L., and coaligned by. X.Y.L and M.Y.W.  Neutron scattering experiments were carried out
by X.Y.L, T.G.P., and P.D.  The data are analyzed by X.Y.L.
The paper was written by P.D., M.W., X.Y.L., C.L.Z. with input from T.A.M, K.H., and G.K..  All coauthors provided comments on the paper.
\end{flushleft}

\begin{flushleft}
{\bf Additional information}
The authors declare no competing financial interests.  Correspondence and
requests for materials should be addressed to P.D., pdai@utk.edu
\end{flushleft}

% ****** Start of file template.aps ****** %
%%
%%
%%   This file is part of the APS files in the REVTeX 4 distribution.
%%   Version 4.0 of REVTeX, August 2001
%%
%%
%%   Copyright (c) 2001 The American Physical Society.
%%
%%   See the REVTeX 4 README file for restrictions and more information.
%%
%
% This is a template for producing manuscripts for use with REVTEX 4.0
% Copy this file to another name and then work on that file.
% That way, you always have this original template file to use.
%
% Group addresses by affiliation; use superscriptaddress for long
% author lists, or if there are many overlapping affiliations.
% For Phys. Rev. appearance, change preprint to twocolumn.
% Choose pra, prb, prc, prd, pre, prl, prstab, or rmp for journal
%  Add 'draft' option to mark overfull boxes with black boxes
%  Add 'showpacs' option to make PACS codes appear
%  Add 'showkeys' option to make keywords appear
%\documentclass[aps,prl,onecolumn,superscriptaddress,amsmath,amssymb]{revtex4}

%\renewcommand{\sectionname}{SSection}
%\renewcommand{\tablename}{Stable}

%\documentclass[aps,prl,preprint,superscriptaddress]{revtex4}
%\documentclass[aps,prl,twocolumn,groupedaddress]{revtex4}

% You should use BibTeX and apsrev.bst for references
% Choosing a journal automatically selects the correct APS
% BibTeX style file (bst file), so only uncomment the line
% below if necessary.
%\bibliographystyle{apsrev}
%\begin{document}
\clearpage
\renewcommand{\figurename}{SFig}
%\preprint{}
\maketitle{\textbf{Supplementary Information: A magnetic origin for high-temperature superconductivity in iron pnictides}}
\\[5mm]
\maketitle{Meng Wang$^\ast$, Chenglin Zhang$^\ast$, Xingye Lu$^\ast$, Guotai Tan, Huiqian Luo, Yu Song, Miaoyin Wang, Xiaotian Zhang, E. A. Goremychkin, T. G. Perring, T. A. Maier, Zhiping Yin, Kristjan Haule, Gabriel Kotliar, Pengcheng Dai}

\author{Meng Wang$^\ast$}
\author{Chenglin Zhang$^\ast$}
\author{Xingye Lu$^\ast$}
\author{Guotai Tan}
\author{Huiqian Luo}
\author{Yu Song}
\author{Miaoyin Wang}
\author{Xiaotian Zhang}
\author{E. A. Goremychkin}
\author{T. G. Perring}
\author{T. A. Maier}
\author{Zhiping Yin}
\author{Kristjan Haule}
\author{Gabriel Kotliar}
\author{Pengcheng Dai}

%\maketitle
\section{Additional data and analysis}

\begin{figure}[h]
\includegraphics[scale=.7]{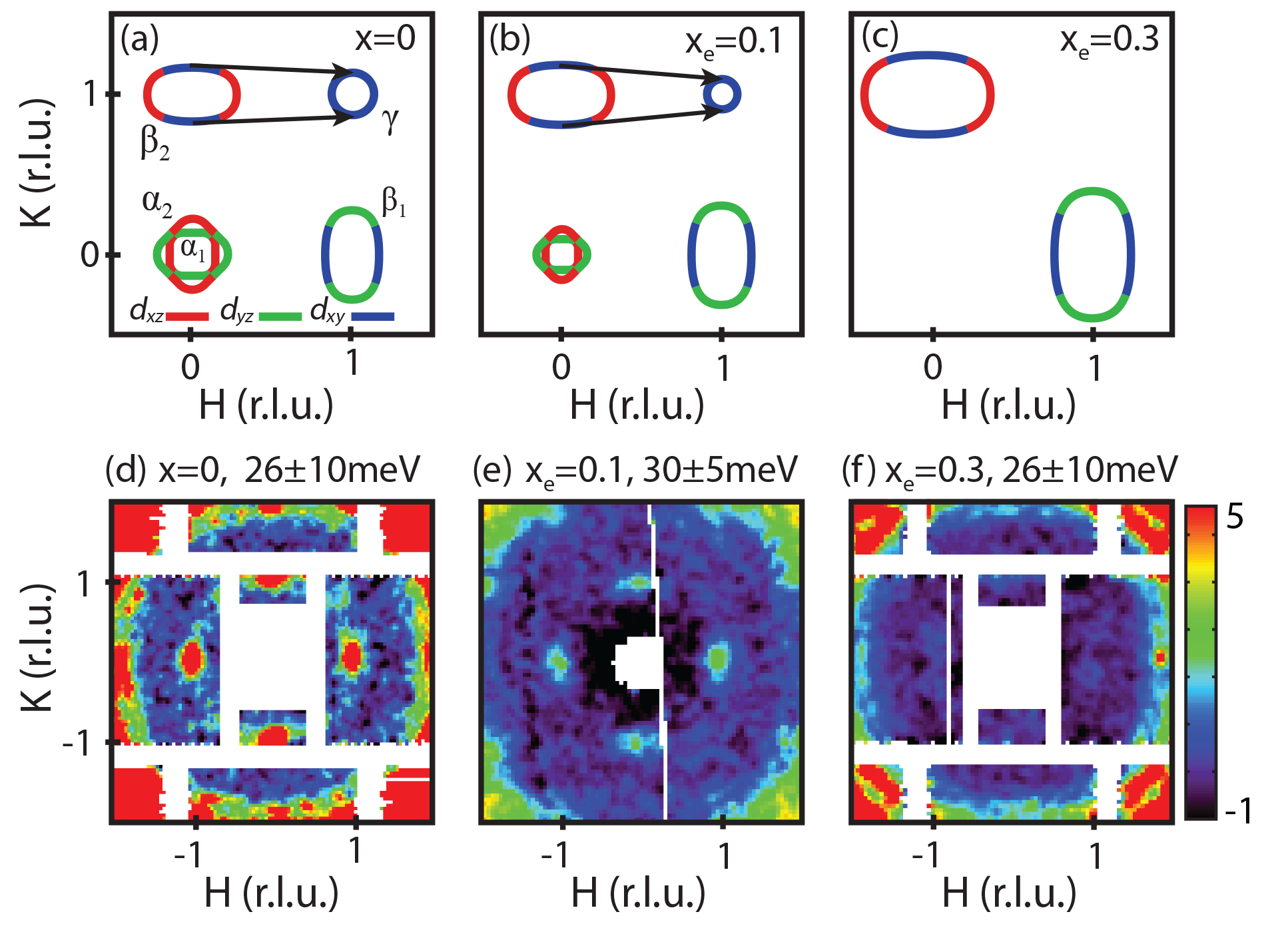}
\caption{
{\bf Schematics of Fermi surface evolution as a function of Ni-doping for BaFe$_{2-x}$Ni$_{x}$As$_2$
and corresponding spin excitations}.
(a,b,c) Evolution of Fermi surfaces with Ni-dopings of $x_e=0,0.1,0.3$. The $d_{xz}$, $d_{yz}$, and $d_{xy}$ orbitals for different Fermi surfaces
are colored as red, green and blue, respectively. (d,e,f) Evolution of low-energy spin excitations for BaFe$_{2-x}$Ni$_{x}$As$_2$
with $x_e=0,0.1,0.3$. For data in (d) and (f), $E_i=450$ meV; (e) $E_i=80$ meV all with $c$-axis along incident beam direction.
 }
\label{SFig0}
\end{figure}

\begin{figure}[h]
\includegraphics[scale=.6]{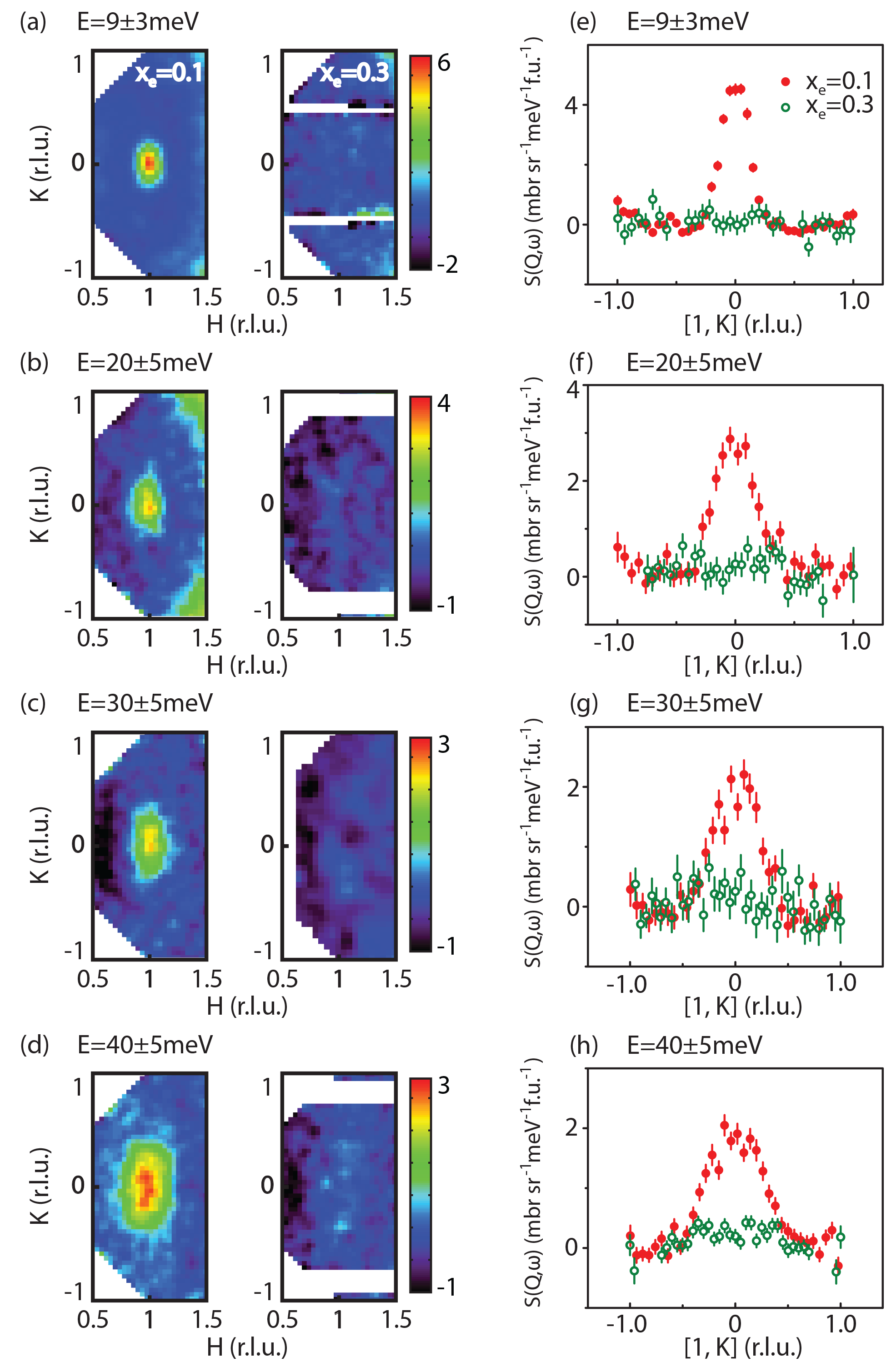}
\caption{ {\bf Comparison of spin excitations in absolute units for BaFe$_{2-x}$Ni$_{x}$As$_2$ with $x_e=0.1,0.3$}.
Evolution of low-energy spin excitations for BaFe$_{2-x}$Ni$_{x}$As$_2$
with $x_e=0.1,0.3$. (a) Data was obtained with $E_i=30$ meV for $x_e=0.1$ and $E_i=80$ meV for
$x_e=0.3$. (b) $E_i=80$ meV for $x_e=0.1$ and $E_i=250$ meV for
$x_e=0.3$. (c) $E_i=80$ meV for $x_e=0.1$ and $E_i=450$ meV for
$x_e=0.3$. (d) $E_i=80$ meV for $x_e=0.1$ and $E_i=250$ meV for
$x_e=0.3$.
(e,f,g,h) The corresponding cuts along the $[1,K]$ direction for these two samples.
 }
\label{SFig1}
\end{figure}

\begin{figure}[h]
\includegraphics[scale=.5]{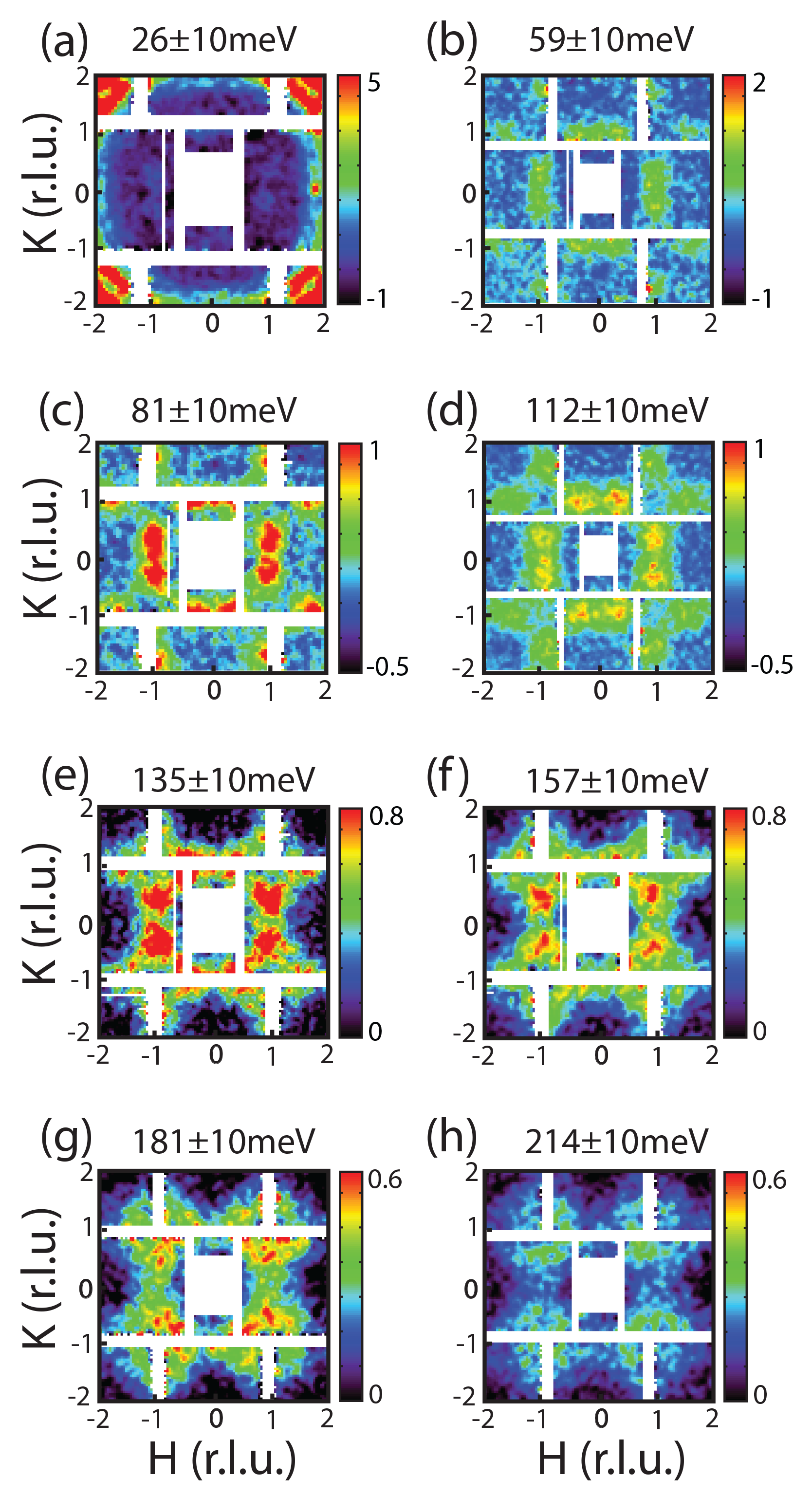}
\caption{ {\bf Spin excitations in absolute units for BaFe$_{1.7}$Ni$_{0.3}$As$_2$}.
(a-h) Evolution of spin excitations for BaFe$_{1.7}$Ni$_{0.3}$As$_2$ all the way to the zone boundary.  Data taken on MAPS at 5 K with $E_i=250$ meV for (b,d) and $E_i=450$ meV for (a,c,e,f,g,h).
 }
\label{SFig2}
\end{figure}

\begin{figure}[h]
\includegraphics[scale=.5]{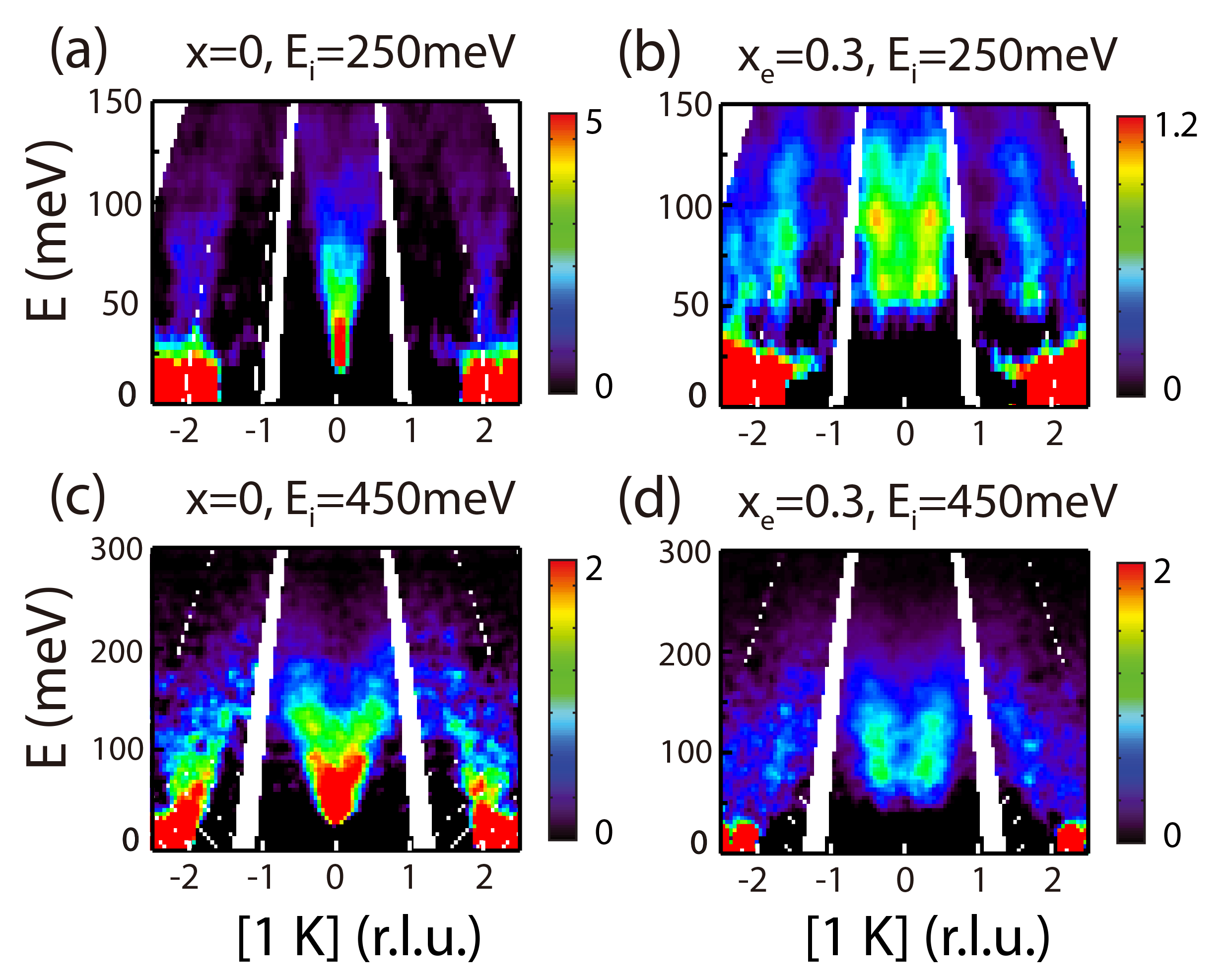}
\caption{ {\bf The dispersion of spin excitations for BaFe$_{2-x}$Ni$_{x}$As$_2$ along $[1, K]$ direction}.
(a, c) The dispersion cuts of BaFe$_2$As$_2$ ($x=0$) with $E_i=250$ meV and $E_i=450$ meV along $[1, K]$ direction. The data is from MAPS. (b, d) Identical dispersion cuts of BaFe$_{1.7}$Ni$_{0.3}$As$_2$ ($x_e=0.3$) at MAPS.
 }
\label{SFignew}
\end{figure}

\begin{figure}[h]
\includegraphics[scale=.5]{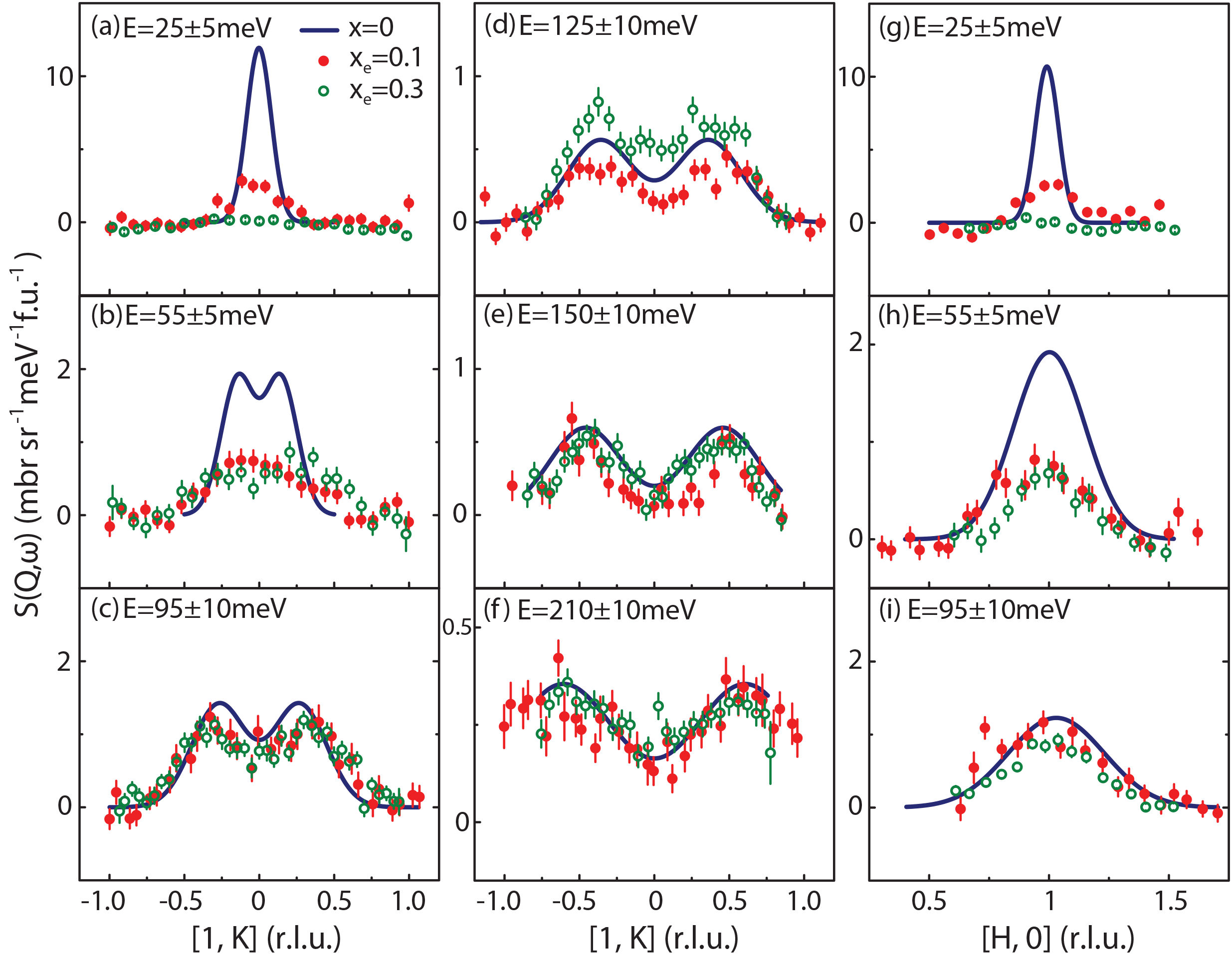}
\caption{ {\bf Spin excitations cuts in absolute units for BaFe$_{2-x}$Ni$_{x}$As$_2$}.
(a-f) Evolution of spin excitations along the $[1,K]$ direction for BaFe$_{2-x}$Ni$_{x}$As$_2$ at $x_e=0,0.1,0.3$.
(g-i) Similar cuts along the $[H,0]$ direction.
 }
\label{SFig3}
\end{figure}

\begin{figure}[h]
\includegraphics[scale=.5]{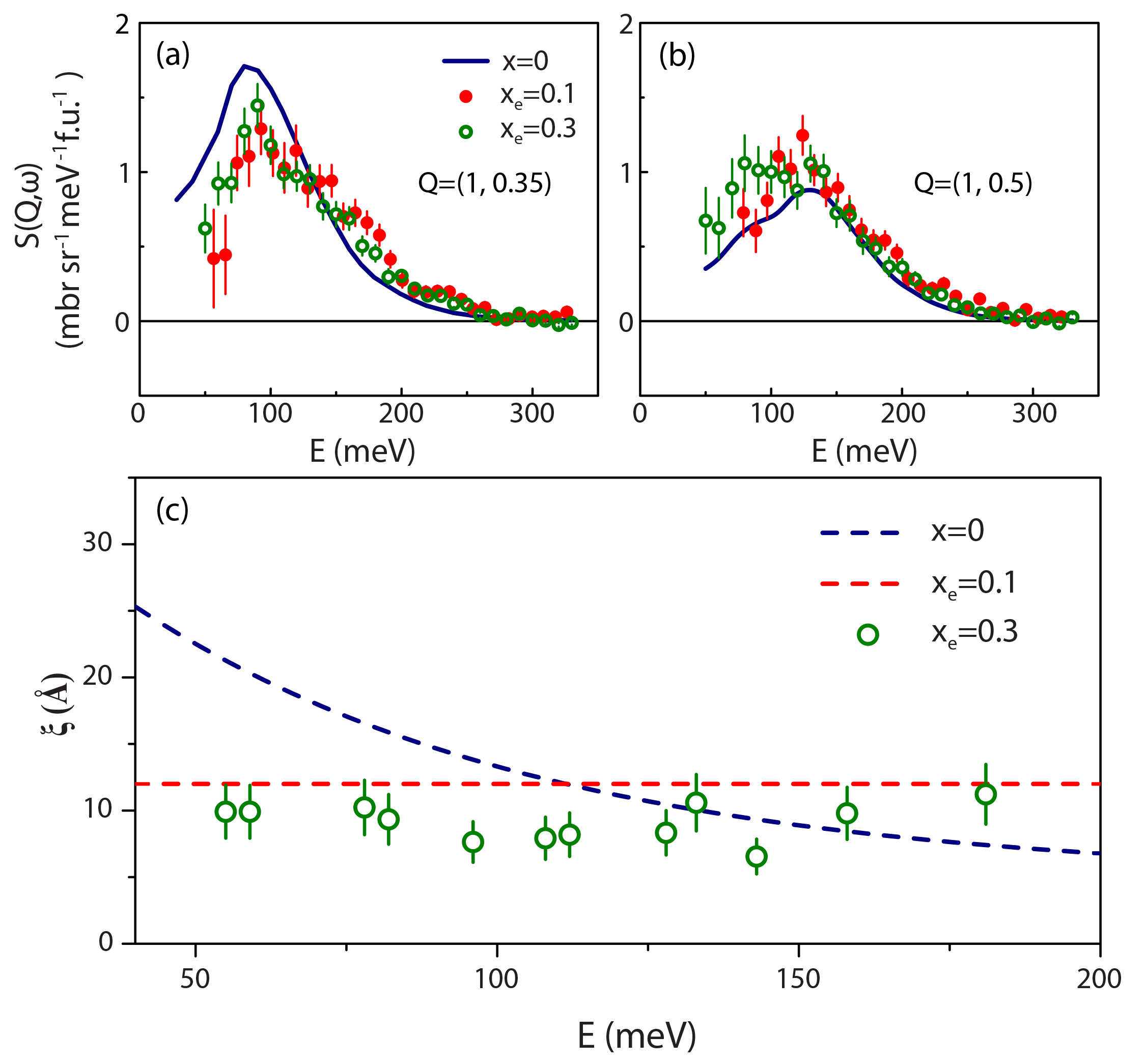}
\caption{ {\bf Spin excitations cuts in absolute units for BaFe$_{2-x}$Ni$_{x}$As$_2$}.
(a,b) Constant-$Q$ cuts of spin excitations for BaFe$_{2-x}$Ni$_{x}$As$_2$ at $x_e=0,0.1,0.3$.
(c) Coherence lengths of spin excitations for these three samples.
 }
\label{SFig4}
\end{figure}

We first discuss detailed experimental results on electron-doped iron pnictides, focusing on
comparison of electron over-doped nonsuperconducting BaFe$_{1.3}$Ni$_{0.3}$As$_2$ with optimally electron-doped
superconductor BaFe$_{1.9}$Ni$_{0.1}$As$_2$ and antiferromagnetic (AF) BaFe$_2$As$_2$.
SFigure \ref{SFig0} shows the evolution of Fermi surfaces as a function
of increasing Ni-doping
obtained from the tight-binding model of Graser {\it et al.} \cite{graser} and our measured low-energy spin excitations spectra for
BaFe$_2$As$_2$,  BaFe$_{1.9}$Ni$_{0.1}$As$_2$, and BaFe$_{1.3}$Ni$_{0.3}$As$_2$.  The absence of hole Fermi pocket near
the zone center for BaFe$_{1.3}$Ni$_{0.3}$As$_2$ means quasiparticle excitations between the hole and electron pockets are not possible, thus eliminating low-energy spin
excitations.  SFigure \ref{SFig1} shows detailed comparison of spin excitations at different energies for BaFe$_{1.9}$Ni$_{0.1}$As$_2$ and
BaFe$_{1.3}$Ni$_{0.3}$As$_2$.   SFigure \ref{SFig2} plots the evolution of spin excitations for BaFe$_{1.3}$Ni$_{0.3}$As$_2$ in
several Brillouin zones. SFigure \ref{SFignew} shows the identical
wave vector versus energy cuts for
 BaFe$_2$As$_2$ \cite{harriger} and BaFe$_{1.7}$Ni$_{0.3}$As$_2$.  The spin excitations in BaFe$_{1.7}$Ni$_{0.3}$As$_2$ are clearly
 absent below 50 meV, which is much bigger than the 15 meV single ion spin anisotropy gap in BaFe$_2$As$_2$ \cite{matan}.
SFigure \ref{SFig3} shows the comparison of cuts along two high-symmetry directions for
BaFe$_{2-x}$Ni$_{x}$As$_2$ at $x_e=0,0.1,0.3$. SFigure \ref{SFig4} plots the constant-$Q$ cuts and spin correlation lengths for these three samples.  It is clear that
the zone boundary spin excitations for electron-doped materials remain almost unchanged at least up to $x_e=0.3$.

\begin{figure}[h]
\includegraphics[scale=.7]{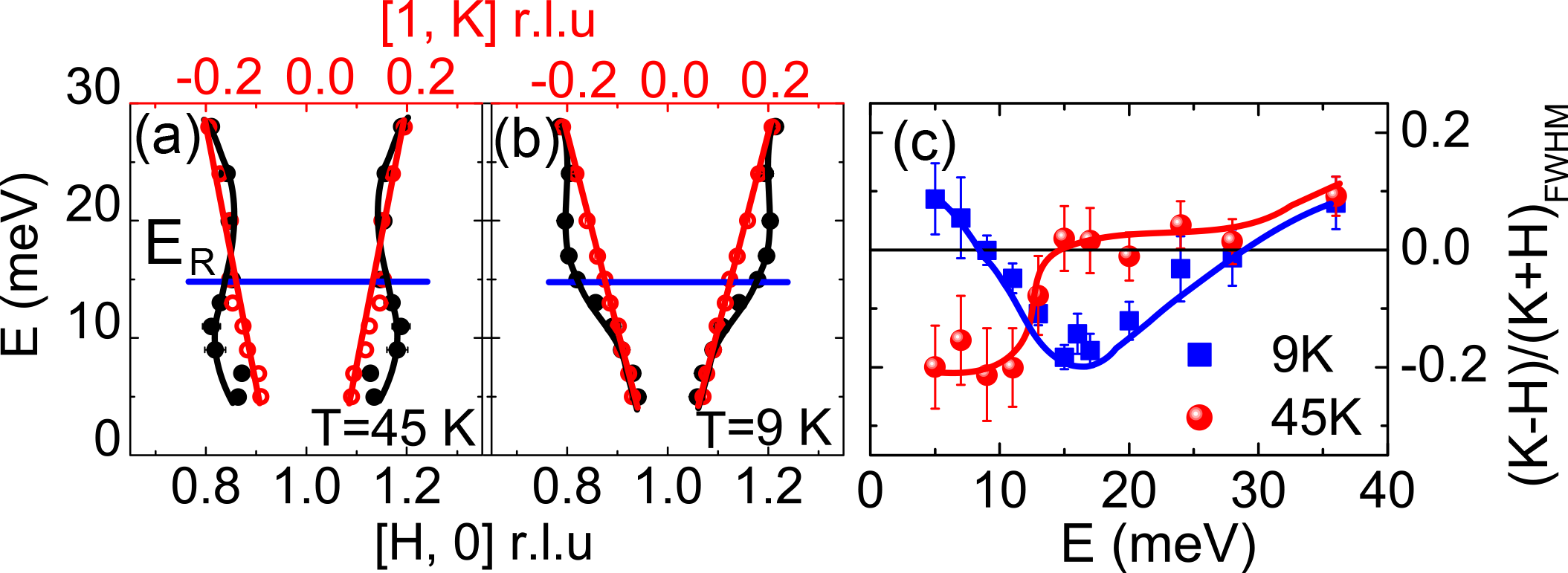}
\caption{
{\bf Dispersion and anisotropy ratio of the low energy spin excitations for Ba$_{0.67}$K$_{0.33}$Fe$_2$As$_2$ above and below $T_c$}.
(a)Normal and (b)superconducting states spin excitation dispersion in full-width at half maximum $(FHWM)$ along the $[H,0]$ and $[1,K]$ directions.
(c) The filled red circles show the anisotropy ratio at $T=45$ K, and the filled blue squares are the same ratio at $9$ K.
The vertical error bars indicate the statistical errors of one standard deviation. The solid lines are a guide to the eye. }
\label{SFig5}
\end{figure}

In Figure 2 of the main text, we have shown that the widths of spin excitations change with increasing energy above $T_c$
for Ba$_{0.67}$K$_{0.33}$Fe$_2$As$_2$.  The resonance width also changes across $T_c$ (Figure 4 of the main text).
To further illustrate the line shape change between the normal and superconducting states, we need to determine
the width ratio of spin excitations along the two high symmetry directions as a function of increasing energy
above and below $T_c$.  We plot the fitting of full width at half maximum (FWHM) of spin excitations along $H$ and $K$  directions in the normal state in SFig. \ref{SFig5} (a). It is a hour-glass like dispersion along $[H,0]$ direction, but a linear dispersion along $[1,K]$ direction. On cooling below $T_c$, the dispersion changes to a shape of flower vase in SFig. \ref{SFig5} (b). we plot in SFig. \ref{SFig5} (c) the energy dependence of the ratio $(K-H)/(K+H)_{FWHM}$ from 5 to 36 meV.
In the normal state at $T=45$ K, spin excitations at energies below the resonance have an oval shape with an elongated $H$ direction,
thus giving negative anisotropy ratio for energies below about 15 meV.
On moving to the resonance energy at $E=15$ meV, the scattering become isotropic and the isotropic
scattering persist up to 30 meV.  For energies above 30 meV, transverse elongated scattering take over, giving positive anisotropy
ratio.  In our previous work \cite{chenglinzhang}, we reported that spin excitations integrated from 10 to 18 meV display an longitudinally
elongated oval shape in the normal state.  This is consistent with present work, which has much better energy resolution and better statistics for measured
spin excitations.  On cooling to the superconducting state,
scattering are essentially isotropic below the resonance, and change to
 the maximum anisotropy at the resonance energy.  For energies above the resonance, the scattering change back to transverse
elongated spin excitations above $35$ meV.

\begin{figure}[h]
\includegraphics[scale=.65]{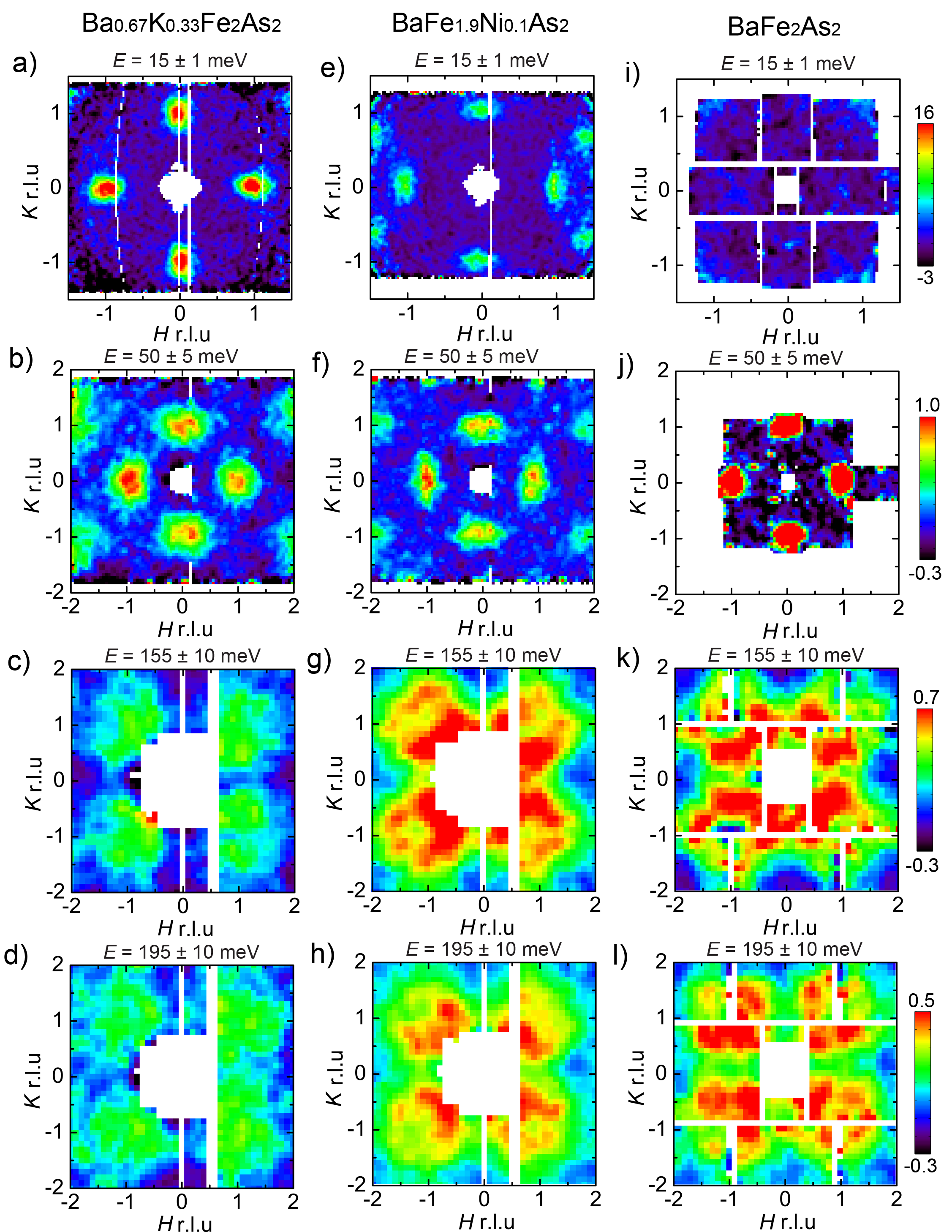}
\caption{
{\bf A comparison of constant-energy images of spin excitations for Ba$_{0.67}$K$_{0.33}$Fe$_2$As$_2$,
BaFe$_{1.9}$Ni$_{0.1}$As$_{2}$, and BaFe$_2$As$_2$ as a function of increasing energy at low-temperature}.
The color bars represent the vanadium normalized absolute spin excitation intensity in the units of
mbarn/sr/meV/f.u. (a) $E=15\pm 1$, (b) $E=50\pm 5$, (c) $E=155\pm 10$, and (d) $E=195\pm 10$ meV are for Ba$_{0.67}$K$_{0.33}$Fe$_2$As$_2$ at $9$ K. (e-h) and (i-l) are identical images for BaFe$_{1.9}$Ni$_{0.1}$As$_{2}$ at $5$ K and BaFe$_2$As$_2$ at $7$ K, respectively.
 }
\label{SFig6}
\end{figure}

In the main text, the integrated dynamic (local) susceptibilities for hole-doped Ba$_{0.67}$K$_{0.33}$Fe$_2$As$_2$, electron-doped BaFe$_{1.9}$Ni$_{0.1}$As$_{2}$,
and BaFe$_{1.3}$Ni$_{0.3}$As$_2$ are shown in Fig. 1h.  To compare spin excitations
of Ba$_{0.67}$K$_{0.33}$Fe$_2$As$_2$ with BaFe$_{2-x}$Ni$_{x}$As$_{2}$ at $x_e=0,0.1$, we show in
SFig. \ref{SFig6} two dimensional constant-energy images at different energies in reciprocal space for these materials.
The magnetic scattering intensities for the three compounds at the same energy are normalized to absolute units in the same color scale.
The measurements for Ba$_{0.67}$K$_{0.33}$Fe$_2$As$_2$ and BaFe$_{1.9}$Ni$_{0.1}$As$_{2}$ were carried out on the MERLIN time-of-flight (TOF) chopper spectrometer at the Rutherford-Appleton Laboratory (RAL), UK. The BaFe$_2$As$_2$ measurements were carried out at MAPS, RAL.
The incident beams were set to be parallel to the $c$-axis of the sample in all three experiments.
Although spin excitation energies are coupled to momentum transfers along the $c$-axis in this scattering geometry,
we note that MERLIN and MAPS experiments have almost identical $L$ for spin excitation energy if the incident beam energy is the same.
Therefore, the raw data with the same $E_i$ for Ba$_{0.67}$K$_{0.33}$Fe$_2$As$_2$, BaFe$_{1.9}$Ni$_{0.1}$As$_{2}$, and BaFe$_2$As$_2$ can be compared directly.
For energy transfer of $E=15\pm1$ meV, the incident beam energy is $E_i=35$ meV for Ba$_{0.67}$K$_{0.33}$Fe$_2$As$_2$, $30$ meV for BaFe$_{1.9}$Ni$_{0.1}$As$_{2}$, and $80$ meV for BaFe$_2$As$_2$. To compare the data quantitatively, the scattering
intensity in \ref{SFig6}(a,e,i) have been corrected by the magnetic form factor.
Spin excitations for Ba$_{0.67}$K$_{0.33}$Fe$_2$As$_2$ have the strongest intensity at the resonance energy ($E=15$ meV) and
are rotated $90^\circ$ in reciprocal space from the line-shape for BaFe$_{1.9}$Ni$_{0.1}$As$_{2}$.
Spin waves in BaFe$_2$As$_2$ has a clear anisotropy spin gap below 15 meV.
For spin excitations at $E=50\pm5$ meV in SFig. \ref{SFig6}, the incident beam energy was $E_i=80$ meV.
Here, spin excitations for Ba$_{0.67}$K$_{0.33}$Fe$_2$As$_2$ are transversely elongated but broader in reciprocal space compared to
those of BaFe$_{1.9}$Ni$_{0.1}$As$_{2}$ and BaFe$_2$As$_2$ (SFig. \ref{SFig6}b, \ref{SFig6}f, \ref{SFig6}i).  The scattering intensity is larger than that of
BaFe$_{1.9}$Ni$_{0.1}$As$_{2}$, but smaller than spin waves of BaFe$_2$As$_2$.  However, the integrated dynamic susceptibility of
Ba$_{0.67}$K$_{0.33}$Fe$_2$As$_2$ is similar with that of BaFe$_2$As$_2$, but larger than that of BaFe$_{1.9}$Ni$_{0.1}$As$_{2}$.
At energies above $E>100$ meV,
spin excitations of
Ba$_{0.67}$K$_{0.33}$Fe$_2$As$_2$ are clearly weaker in intensity than that of BaFe$_{1.9}$Ni$_{0.1}$As$_{2}$ and BaFe$_2$As$_2$. The constant-energy images have been subtracted by a radial background for Ba$_{0.67}$K$_{0.33}$Fe$_2$As$_2$ and a constant background for BaFe$_{1.9}$Ni$_{0.1}$As$_{2}$ and BaFe$_2$As$_2$.

\begin{figure}[h]
\includegraphics[scale=.65]{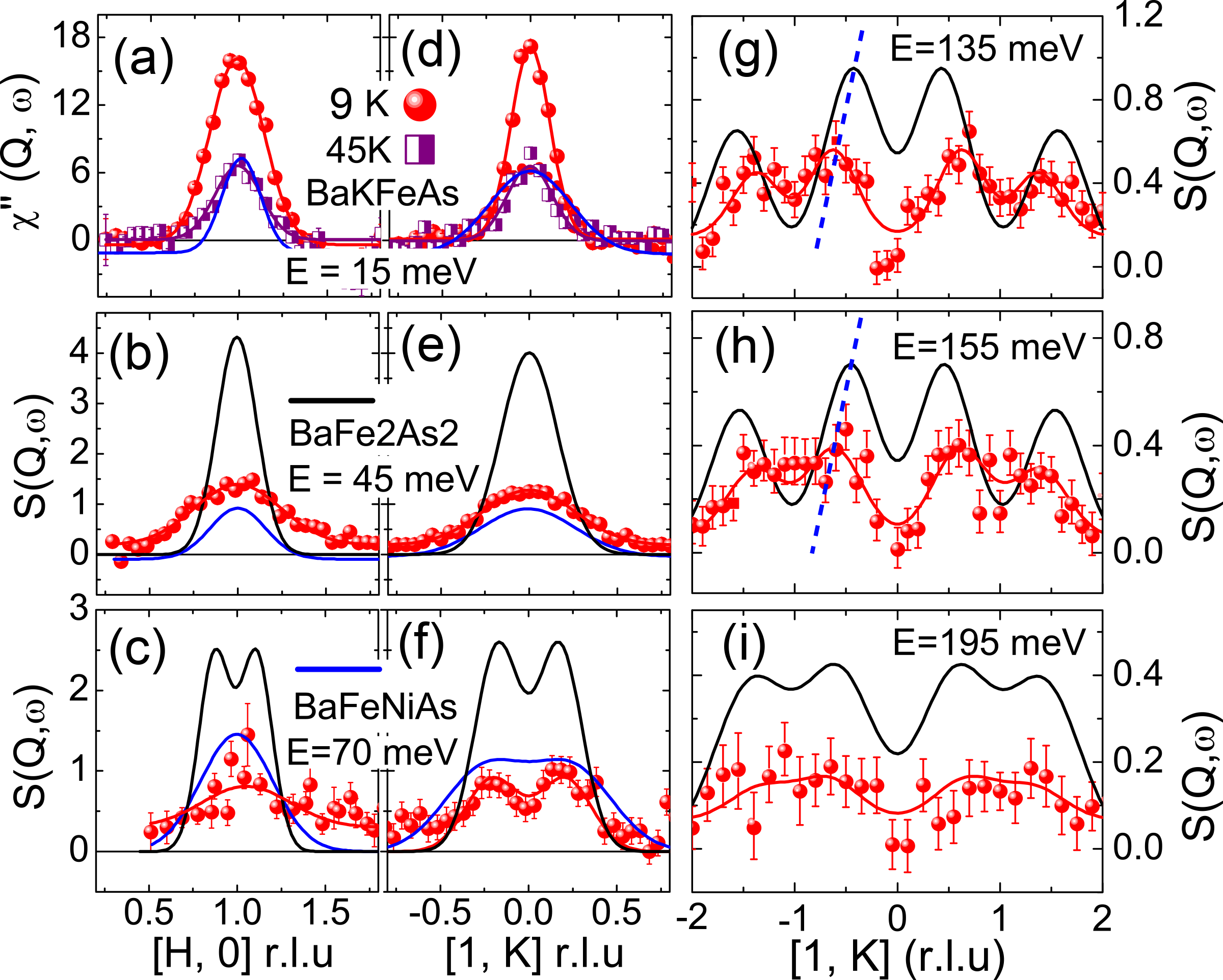}
\caption{
{\bf Constant-energy cuts of spin excitations as a function of increasing energy}.
(a-c) are the cuts along the $[H, 0]$ direction and (d-i) are along the $[1, K]$ direction for Ba$_{0.67}$K$_{0.33}$Fe$_2$As$_2$ at $9$ K (red filled circles), $45$ K (half filled purple squares), BaFe$_{1.9}$Ni$_{0.1}$As$_{2}$ at $5$ K (solid blue line) and BaFe$_2$As$_2$ at $7$ K (solid black line). (a) $E=15\pm1$ meV with $E_i=35$ meV for Ba$_{0.67}$K$_{0.33}$Fe$_2$As$_2$ and $E_i=30$ for BaFe$_{1.9}$Ni$_{0.1}$As$_{2}$, (b) $E=45\pm5$ meV with Ei=80meV, (c) $E=70\pm5$ meV with $E_i=250$ meV are cuts along $[H, 0]$ direction. (d-f) are the same constant-energy cuts along $[1, K]$ direction. (g) $E=135\pm10$, (h) $E=155\pm10$, (i) $E=195\pm10$ meV are from $E_i=450$ meV data. The solid lines are Gaussian fits to the data. (a,d) have been corrected the Bose population factor and magnetic form factor.
}
\label{SFig7}
\end{figure}

To quantitatively determine the dispersions of spin excitations
in Ba$_{0.67}$K$_{0.33}$Fe$_2$As$_2$, BaFe$_{1.9}$Ni$_{0.1}$As$_{2}$, and BaFe$_2$As$_2$,
we cut through $[H, 0]$ and $[1, K]$ directions of the two dimensional scattering images in SFig. \ref{SFig6}.
SFig. \ref{SFig7}a-\ref{SFig7}c  and \ref{SFig7}d-\ref{SFig7}e show constant-energy cuts at energies of $E=15\pm1$, $45\pm5$, $70\pm5$ meV along the $[H, 0]$ and $[1, K]$ directions for Ba$_{0.67}$K$_{0.33}$Fe$_2$As$_2$, BaFe$_{1.9}$Ni$_{0.1}$As$_{2}$, and BaFe$_2$As$_2$, respectively.  SFig. \ref{SFig7}a,\ref{SFig7}d cuts have been corrected the effect of Bose population factor and magnetic form factor, since they have different incident beam energies and $T=45$ K data.
The cuts along the $[1, K]$ direction at $E=135\pm10$, $155\pm10$ and $195\pm10$ meV reveal the weaker susceptibility at high energies for Ba$_{0.67}$K$_{0.33}$Fe$_2$As$_2$. The dashed blue lines in SFig. \ref{SFig7}g, \ref{SFig7}h indicate that
spin excitations of Ba$_{0.67}$K$_{0.33}$Fe$_2$As$_2$ disperse more rapidly and have a softened band top.

\begin{figure}[h]
\includegraphics[scale=.5]{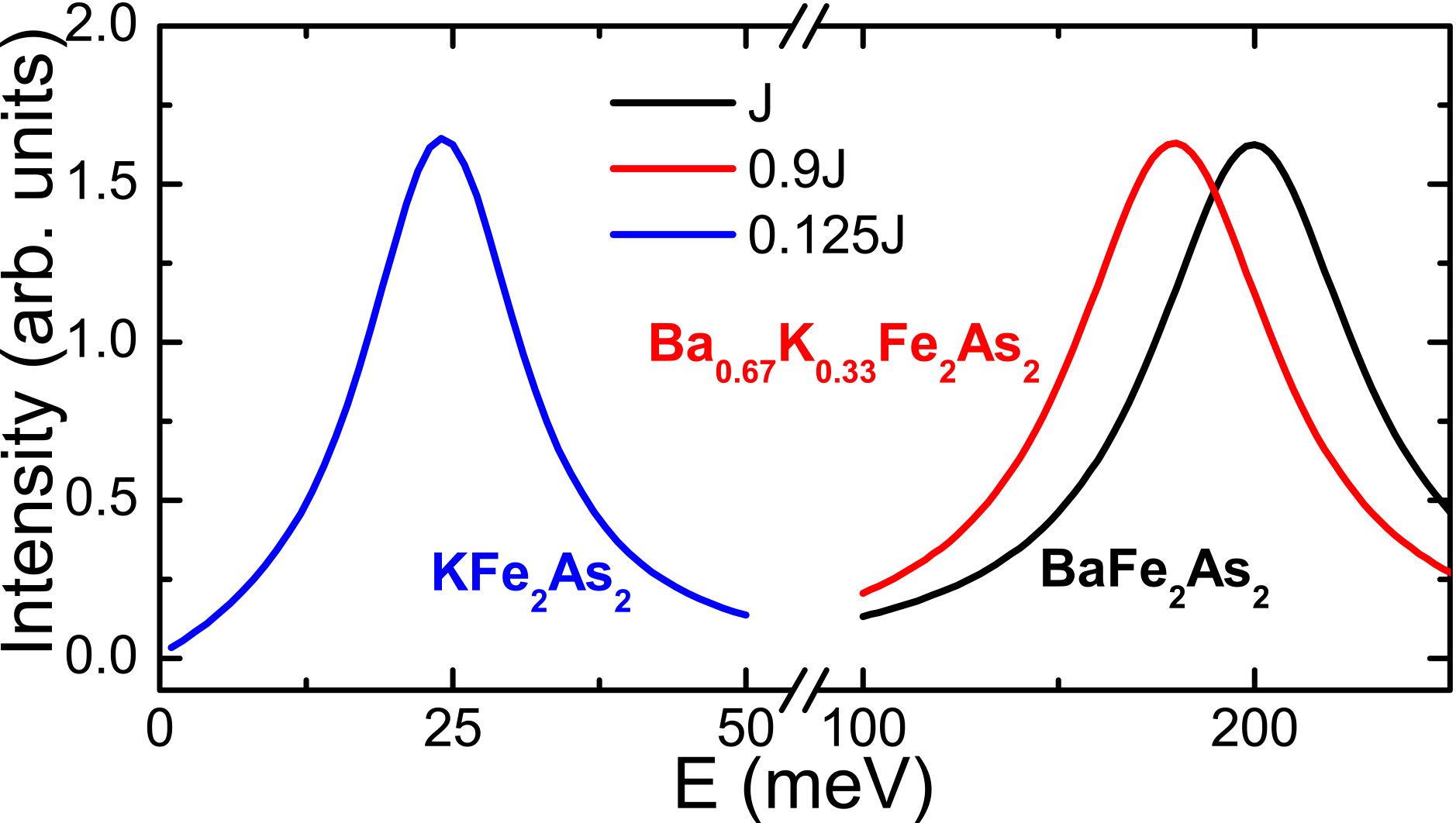}
\caption{
{\bf The effect of magnetic exchange couplings $J$ ($J_{1a}$, $J_{1b}$ and $J_2$) on the band top of spin excitations}.
The black line is energy cut at ($0.8$ \textless $H$ \textless $1.2$, $0.8$ \textless $H$ \textless $1.2$) r.l.u for BaFe$_2$As$_2$ in the Heisenberg spin wave model \cite{harriger}.  The red line is for Ba$_{0.67}$K$_{0.33}$Fe$_2$As$_2$ with $10$ \% soften band top. The blue line is a similar estimation
for KFe$_2$As$_2$ assuming zone boundary is around $E=25$ meV.
 }
 \label{SFig8}
\end{figure}

\begin{figure}[h]
\includegraphics[scale=.9]{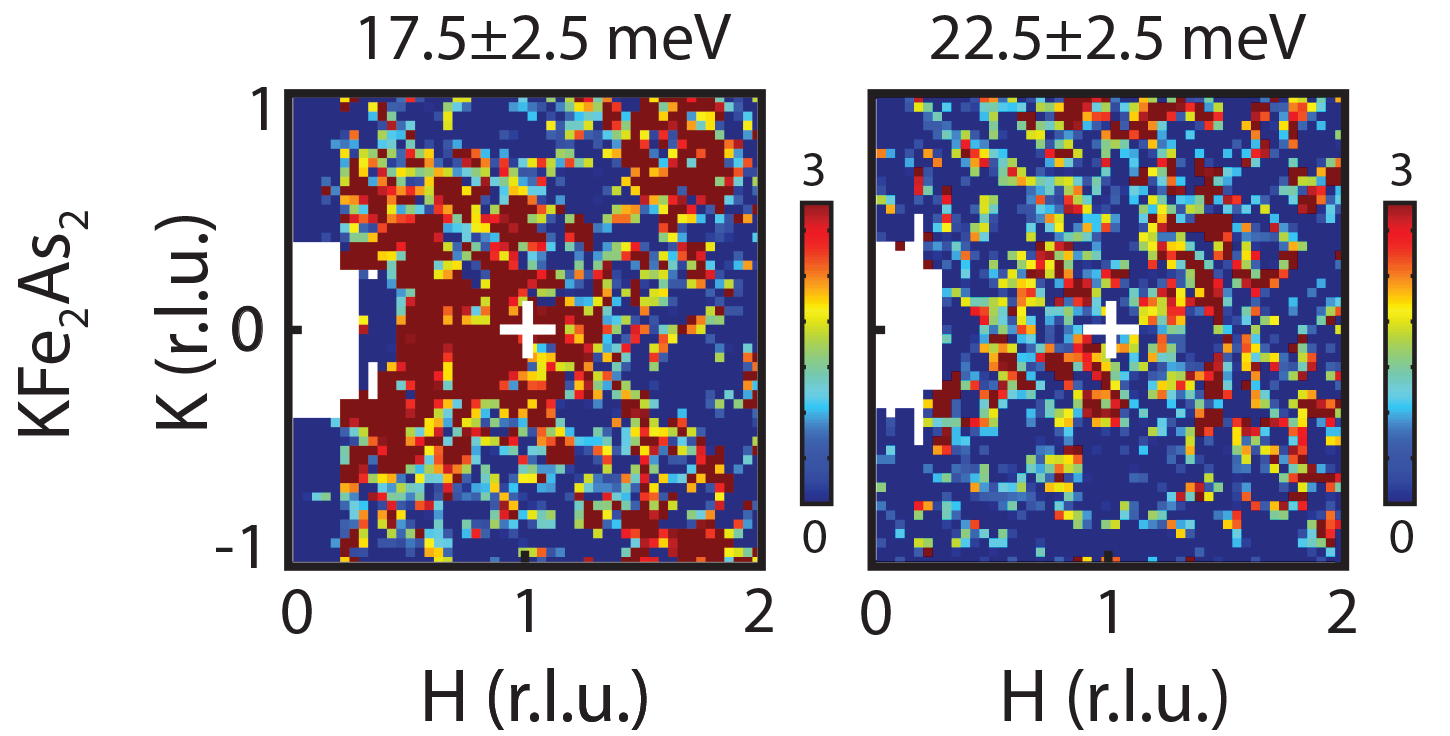}
\caption{
{\bf The disappearance of sin excitations above 25 meV for KFe$_2$As$_2$}.
Using $E_i=80$ meV, we can still see clear incommensurate spin excitations at
(a) $E=17.5\pm2.5$ meV.  However, for energy transfers
at $E=22.5\pm2.5$ mV, one can no longer see spin excitations, indicating
a strong suppression of the spin excitation spectral weight for $E>25$ meV.
 }
 \label{SFig9}
\end{figure}

Spin excitations disappear at the zone boundary $[1, 1]$ in Ba$_{0.67}$K$_{0.33}$Fe$_2$As$_2$,
BaFe$_{1.3}$Ni$_{0.3}$As$_2$,
 and BaFe$_2$As$_2$.
The band top is governed by the effective  magnetic exchange couplings $J$ ($J_{1a}$, $J_{1b}$ and $J_2$), as defined in BaFe$_2$As$_2$ \cite{harriger}. To estimate the change of $J$ for hole-doped Ba$_{0.67}$K$_{0.33}$Fe$_2$As$_2$, we calculate the energy cut at $[1, 1]$ by exploring the Heisenberg Hamiltonian of parent compound. It turns out that $J_{1a}$, $J_{1b}$ and $J_2$ have comparable effect on the band top. Based on the dispersion of Ba$_{0.67}$K$_{0.33}$Fe$_2$As$_2$, the effective magnetic exchange $J$ is found to be about $10$ \% smaller for Ba$_{0.67}$K$_{0.33}$Fe$_2$As$_2$ compared
with that of BaFe$_2$As$_2$ (SFig. \ref{SFig8}).  For comparison, if we assume the band top for KFe$_2$As$_2$ is around $E=25$ meV, the
effective magnetic exchange should be about 90\% smaller for KFe$_2$As$_2$. Of course, we know this is not an accurate estimation
since spin excitations in KFe$_2$As$_2$ are incommensurate and have an inverse dispersion.  In any case, given the zone boundary energy of $E\approx 25$ meV,
the  effective magnetic exchange couplings in KFe$_2$As$_2$ must be much smaller than that of BaFe$_2$As$_2$.
SFigure \ref{SFig9} shows additional data for KFe$_2$As$_2$ that clearly reveal a dramatic reduction in magnetic scattering above $E=20$ meV.

\section{Random Phase Approximation (RPA) and DFT+DMFT calculations}

\begin{figure}[h]
\includegraphics[scale=.5]{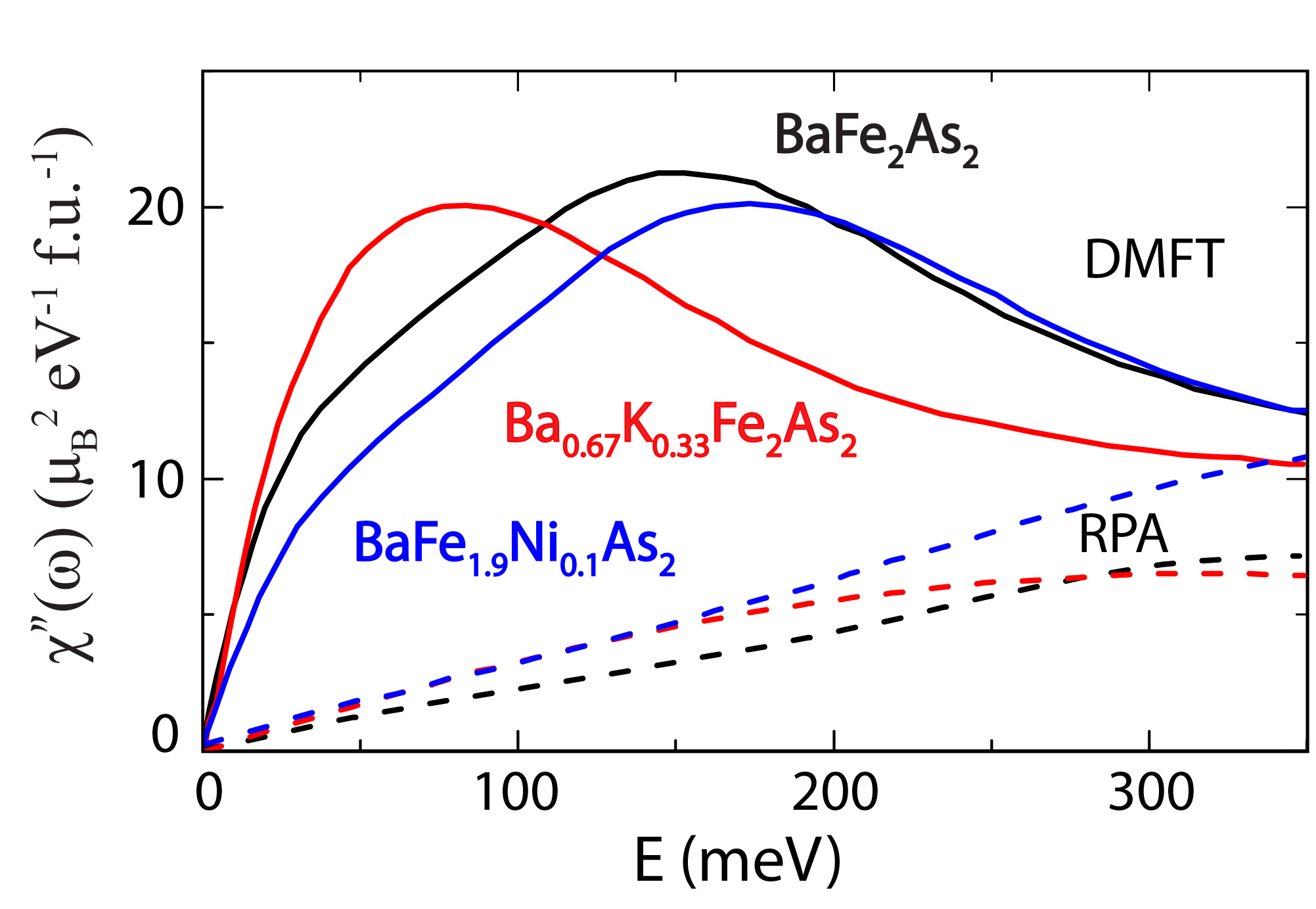}
\caption{
{\bf RPA and LDA+DMFT calculated local susceptibility for different iron pnictides}.
RPA and LDA+DMFT calculations of $\chi^{\prime\prime}(\omega)$ in absolute units for
 Ba$_{0.67}$K$_{0.33}$Fe$_2$As$_2$ comparing with earlier results for
 BaFe$_2$As$_2$ and BaFe$_{1.9}$Ni$_{0.1}$As$_2$ \cite{mliu}.
 }
 \label{SFig11}
\end{figure}

As discussed in the main text, RPA calculation of
 the wave vector dependence of spin excitations in hole-doped Ba$_{0.67}$K$_{0.33}$Fe$_2$As$_2$
is in clear disagreement with experiments (Fig. 3 of the main text).    Although
a combined DFT and DMFT approach \cite{mliu,park} still does not agree in detail
with the data (Fig. 3), it captures the trend of spectral weight transfer away from
$Q_{AF}=(1,0)$ on increasing the energy and forming a pocket centered at $Q=(1,1)$.
The solid red, black, and blue lines in SFigure \ref{SFig11} show calculated local susceptibility
in absolute units based on a combined DFT and DMFT approach for Ba$_{0.67}$K$_{0.33}$Fe$_2$As$_2$, BaFe$_2$As$_2$, and BaFe$_{1.9}$Ni$_{0.1}$As$_{2}$ \cite{mliu}, respectively.
This theoretical method predicts that electron
doping to BaFe$_2$As$_2$ does not affect the spin susceptibility at high
energy ($E>150$ meV), while spin excitations in the hole doped compound
beyond 100 meV are suppressed by shifting the spectral weight to
lower energies.  This is in qualitative agreement with our absolute intensity measurements (Fig. 1h).
The reduction of the high energy spin spectral weight and its transfer to low energy with hole doping, but not with electron doping, is not naturally explained by the band theory,  and requires models which incorporate both the itinerant quasiparticles and the local moment physics.
The hole doping makes electronic state more correlated, as local moment formation is strongest in the half-filled $d^5$ shell,
and mass enhancement larger thereby reducing the electronic energy scale in the problem.

Our theoretical DFT+DMFT method for computing the magnetic excitation spectrum
employs the abinitio full potential implementation of the method, as detailed in \cite{Haule}.
The DFT part is based on the density functional theory (DFT) code of
Wien2k \cite{Blaha}.  The DMFT method requires solution of the generalized
quantum impurity problem, which is here solved by the numerically
exact continuous-time quantum Monte Carlo method \cite{Haule07,werner}. The Coulomb
interaction matrix for electrons on iron atom was determined by the
self-consistent GW method in Ref. \cite{kutepov}, giving $U = 5$ eV and $J = 0.8$ eV
for the local basis functions within the all electron approach
employed in our DFT+DMFT method.  The dynamical magnetic
susceptibility $\chi^{\prime\prime}(Q,E)$ is computed from the $ab$ initio perspective by
solving the Bethe-Salpeter equation, which involves the fully
interacting one particle Greens function computed by DFT+DMFT, and the
two particle vertex, also computed within the same method (for details
see Ref. \cite{park}).  We computed the two-particle irreducible vertex
functions of the DMFT impurity model, which coincides with the local
two-particle irreducible vertex within DFT+DMFT method. The latter is
assumed to be local in the same basis in which the DMFT self-energy is
local, here implemented by projection to the muffin-tin sphere.

\section{Magnetic Exchange Energy and superconducting condensation energy for Ba$_{0.67}$K$_{0.33}$Fe$_2$As$_2$}

In a neutron scattering experiment, we measure scattering function $S(q,E=\hbar\omega)$ which is related to the imaginary part of the dynamic
susceptibility via $S(q,\omega)=[1+n(\omega,T)]\chi^{\prime\prime}(\vec{q},\omega)$, where  $[1+n(\omega,T)]$ is be Bose population factor.
The magnetic exchange coupling and the imaginary part of spin susceptibility are related via the formula \cite{scalapino}:
\begin{equation}\begin{split}
\langle \vec{S}_i \cdot \vec{S}_{j}\rangle&=\frac{3}{\pi g^2\mu_{B}^{2}}\int \frac{d\vec{q}^2}{(2\pi)^2}\int d\omega [1+n(\omega,T)]\chi^{\prime\prime}(\vec{q},\omega) cos[\vec{q}\cdot(\vec{i}-\vec{j})],\\
\end{split}\end{equation}
where $g=2$ is the Land$\rm \acute{e}$ $g$-factor. The magnetic exchange energy can be written as
\begin{equation}\begin{split}
E_{ex}=&\sum_{<i,j>}J_{ij}\langle \vec{S}_i \cdot \vec{S}_{j}\rangle\\
      =&\frac{3}{\pi g^2\mu_{B}^{2}}\int \frac{d\vec{q}^2}{(2\pi)^2}\int d\omega\\
     \{& \sum_{i}J_{1a} [1+n(\omega,T)] \chi^{\prime\prime}(\vec{q},\omega) cos(q_x)\\
     +&\sum_{i}J_{1b} [1+n(\omega,T)] \chi^{\prime\prime}(\vec{q},\omega) cos(q_y)\\
     +&\sum_{i}J_{2} [1+n(\omega,T)] \chi^{\prime\prime}(\vec{q},\omega) [cos(q_x+q_y)+cos(q_x-q_y)]\}.\\
\end{split}\end{equation}
Here we have assumed an anisotropic model for the effective magnetic exchange coupling \cite{harriger}, different from the case of
copper oxide superconductors \cite{scalapino}.  $J_{1a}$ is the effective magnetic coupling strength between two nearest sites along the $a$ direction, while $J_{1b}$ is that along the $b$ direction, and $J_{2}$ is the coupling between the next nearest neighbor sites.  Hence we are able to obtain the change in magnetic exchange energy between the superconducting and normal states by the experimental data of $\chi^{\prime\prime}(\vec{q},\omega)$ in both states.
Strictly speaking, we want to estimate the zero temperature
difference of the magnetic exchange energy between the normal
and the superconducting states, and use the outcome to compare
with the superconducting condensation energy \cite{scalapino}.
Unfortunately, we do not have direct information on the normal state $\chi^{\prime\prime}(\vec{q},\omega)$ at zero temperature. Nevertheless, since our neutron scattering measurements at low-energies
showed that the $\chi^{\prime\prime}(\vec{q},\omega)$ are very similar below and above $T_c$ near the AF wave vector $Q_{AF}=(1,0,1)$ and only a very shallow spin gap at
$Q=(1,0,0)$ (see Figs. 1f and 1h in \cite{chenglinzhang}),
we assume that there are negligible changes in $\chi^{\prime\prime}(\vec{q},\omega)$ above and below $T_c$ at zero temperature for energies below 5 meV.  For spin excitation energies above 6 meV,
Bose population factors between 7 K and 45 K  are negligibly small.
In previous work on optimally doped YBa$_2$Cu$_3$O$_{6.95}$ superconductor, we have assumed that spin excitations in the normal state at zero temperature are
negligibly small and thus do not contribute to the exchange energy \cite{woo}.

The directly measured quantity is the scattering differential cross section
\begin{equation}\begin{split}
\frac{d^2\sigma}{d\Omega dE}\frac{k_i}{k_f}&=\frac{2(\gamma r_e)^2}{\pi g^2\mu_{B}^{2}}|F(\vec{Q})|^{2} [1+n(\omega,T)] \chi^{\prime\prime}(\vec{q},\omega),\\
\end{split}\end{equation}
where $k_{i}$ and $k_{f}$ are the magnitudes of initial and final neutron momentum and $F(\vec{Q})$ is the Fe magnetic form factor, and $(\gamma r_e)^2=0.2905\ {\rm barn\cdot sr^{-1}}$.

\begin{figure}[H]
\begin{center}
\includegraphics[scale=0.8]{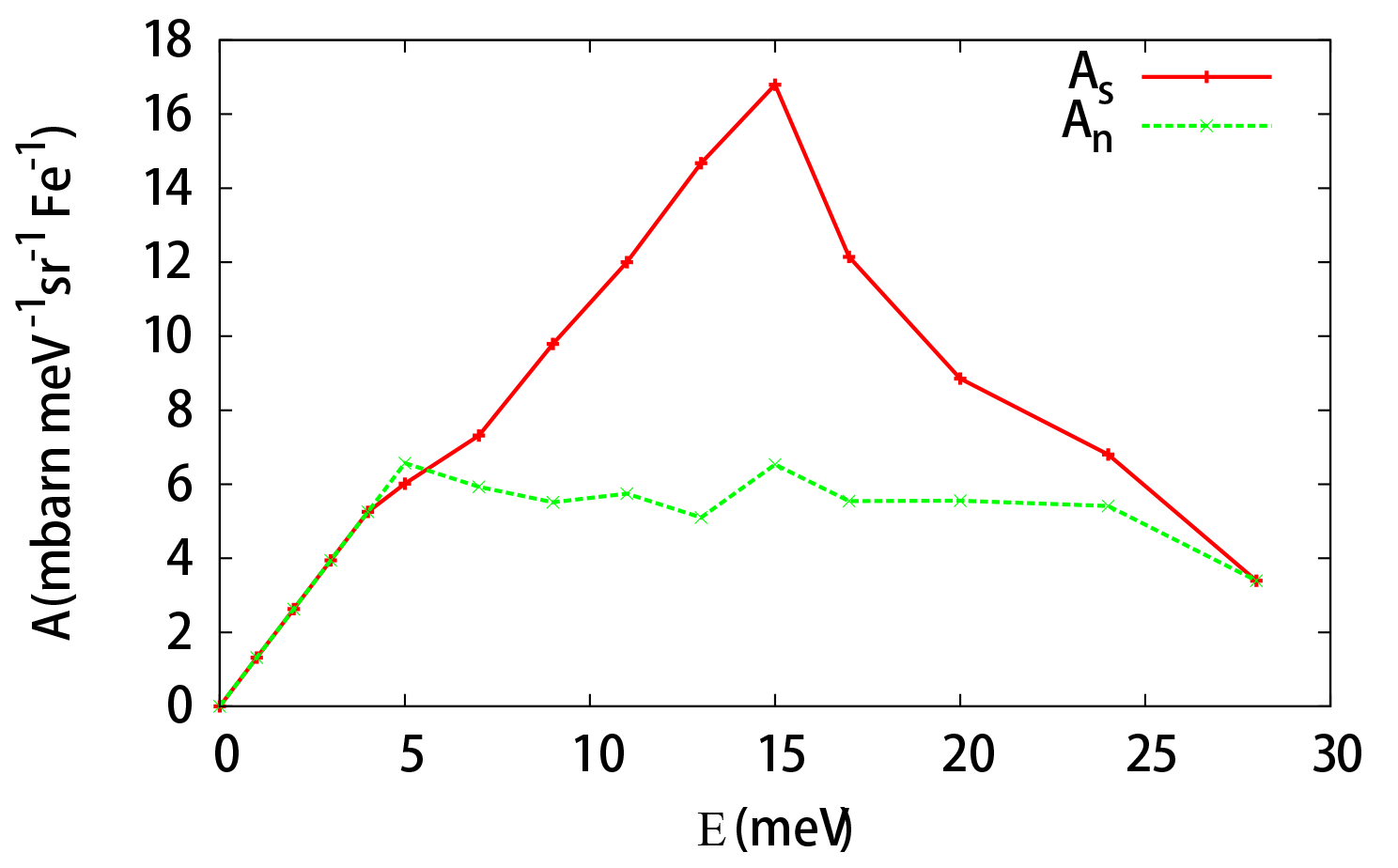}
\caption{
{ Assumed $A_{s(n)}$ below 5 meV.}
}\label{a}
\end{center}
\label{SFig12}
\end{figure}

The quantity $\frac{2(\gamma r_e)^2}{\pi g^2\mu_{B}^{2}}\chi^{\prime\prime}(\vec{q},E)$ in both superconducting and normal states can be fitted by a Gaussian $A_{s(n)}e^{-[\frac{(q_x-1)^2}{2\sigma_{x,s(n)}^{2}}+\frac{q_y^2}{2\sigma_{y,s(n)}^{2}}]}$ for resonance wave vector $(1,0)$ and by cutting the raw data.
The outcome is summarized in the table \ref{data}:

\begin{table}[H]\label{data}
 \centering
 \begin{tabular}{|p{0.9cm}p{1.2cm}p{1.2cm}p{1.2cm}p{1.2cm}p{1.2cm}p{1.2cm}|}\hline
$E$       &$\sigma_{x,s}$ &$\sigma_{y,s}$ &$A_s$      &$\sigma_{x,n}$ &$\sigma_{y,n}$ &$A_n$\\
\hline
$5$ &$0.050$    &$0.060$    &$6.017$     &$0.115$   &$0.076$    &$6.571$ \\
\hline
$7$ &$0.059$    &$0.066$    &$7.318$     &$0.109$   &$0.080$    &$5.929$\\
\hline
$9$ &$0.077$    &$0.077$    &$9.789$     &$0.154$   &$0.100$    &$5.515$\\
\hline
$11$    &$0.092$    &$0.083$    &$12.001$   &$0.160$    &$0.107$    &$5.745$\\
\hline
$13$    &$0.121$    &$0.097$    &$14.674$   &$0.145$    &$0.124$    &$5.102$\\
\hline
$15$    &$0.153$    &$0.106$    &$16.792$   &$0.125$    &$0.130$    &$6.535$\\
\hline
$17$    &$0.167$    &$0.118$    &$12.141$   &$0.152$    &$0.157$    &$5.546$\\
\hline
$20$    &$0.173$    &$0.136$    &$8.856$     &$0.131$   &$0.128$    &$5.555$\\
\hline
$24$    &$0.165$    &$0.155$    &$6.802$     &$0.134$   &$0.146$    &$5.411$\\
\hline
$28$    &$0.182$    &$0.177$    &$3.393$     &$0.161$   &$0.165$    &$3.393$\\
\hline

 \end{tabular}
 \label{data}
\end{table}
where the unit of $E$ is meV and that of $A_{s(n)}$ is mbarn$\cdot$meV$^{-1} \cdot$sr$^{-1}\cdot$Fe$^{-1}$. For the case below 5 meV, we assume that $A_n$ decreases to zero linearly with energy and $A_s=A_n$ (see Fig. 1h in Ref. \cite{chenglinzhang}), while the $\sigma$'s keep the values at 5 meV.
The assumption is shown in SFig. \ref{SFig12}, where the resonance is seen at $E=15$ meV.

Because the condensation energy is only defined at zero temperature, we take $T=0$ in equation (1) and the integral gives:
\begin{equation}\begin{split}
\langle \vec{S}_i \cdot \vec{S}_{i+x}\rangle_s-\langle \vec{S}_i \cdot \vec{S}_{i+x}\rangle_n&=-0.0039,\\
\langle \vec{S}_i \cdot \vec{S}_{i+y}\rangle_s-\langle \vec{S}_i \cdot \vec{S}_{i+y}\rangle_n&=0.0043,\\
\sum_{l=x\pm y}(\langle \vec{S}_i \cdot \vec{S}_{i+l}\rangle_s-\langle \vec{S}_i \cdot \vec{S}_{i+l}\rangle_n)&=-0.0073\\
\end{split}\end{equation}

The magnetic exchange coupling constants in an anisotropic model are estimated to be
\begin{equation}\begin{split}
J_{1a}S&=53.3\ {\rm meV},\\
J_{1b}S&=-8.3\ {\rm meV},\\
J_{2}S&=12.2\ {\rm meV},\\
\end{split}\end{equation}
which are $10\%$ smaller than that of BaFe$_2$As$_2$ \cite{harriger} and we estimate $S$ to be close to $\frac{1}{2}$ \cite{mengshu,jun}.  Hence the exchange energy change is
\begin{equation}\begin{split}
\Delta E_{ex}&=-0.66~{\rm meV/Fe.}\\
\end{split}\end{equation}

The condensation energy $U_c$ for optimally doped Ba$_{0.68}$K$_{0.32}$Fe$_2$As$_2$ can be calculated to be
\begin{equation}\begin{split}
          U_c&=-17.3~{\rm J/mol}\\
          &=-17.3~\frac{1\ {\rm eV}}{1.6\times 10^{-19}}\frac{1}{6.02\times 10^{23} {\rm f.u.}} \\
          &=-17.3~\frac{1\ {\rm eV}}{1.6\times 10^{-19}}\frac{1}{2\times 6.02\times 10^{23}\ {\rm Fe}} \\
          &=-0.09~{\rm meV/Fe}
\end{split}\end{equation}
from the specific heat data of Ref. \cite{popovich}.  Therefore, we have the ratio $\Delta E_{ex}/U_c\approx 7.4$, meaning that the change in the magnetic
exchange energy is sufficient to account for the superconducting condensation energy in Ba$_{0.68}$K$_{0.32}$Fe$_2$As$_2$.  We note that a similar calculation for heavy
Fermion superconductor CeCu$_2$Si$_2$ also reveals that the change in
magnetic exchange energy
is sufficient to account for the superconducting condensation energy \cite{stockert}.

We thank M. S. Liu and L. W. Harriger for providing the BaFe$_{1.9}$Ni$_{0.1}$As$_{2}$ and BaFe$_2$As$_2$ raw data, respectively.

%\bibliography{NoEndingPoint}

\end{document}